\documentclass[a4paper,11pt]{article}
\pdfoutput=1

\usepackage{jheppub}

\usepackage[T1]{fontenc}

\usepackage{xspace,units,graphicx}

\newcommand{\numu}{\ensuremath{\nu_{\mu}}\xspace}
\newcommand{\numubar}{\ensuremath{\overline{\nu}_{\mu}}\xspace}
\newcommand{\nue}{\ensuremath{\nu_{e}}\xspace}
\newcommand{\nuebar}{\ensuremath{\overline{\nu}_{e}}\xspace}
\newcommand{\nutau}{\ensuremath{\nu_{\tau}}\xspace}

\title{The MINOS experiment: results and prospects}

\author[a,1]{J.J. Evans,\note{For the MINOS collaboration.}}

\affiliation[a]{University of Manchester,\\Department of Physics and Astronomy, Oxford Road, Manchester, M13 9PL, United Kingdom}

\emailAdd{justin.evans@hep.manchester.ac.uk}

\abstract{ The MINOS experiment has used the world's most powerful
  neutrino beam to make precision neutrino oscillation
  measurements. By observing the disappearance of muon neutrinos,
  MINOS has made the world's most precise measurement of the larger
  neutrino mass splitting, and has measured the neutrino mixing angle
  $\theta_{23}$. Using a dedicated antineutrino beam, MINOS has made
  the first direct precision measurements of the corresponding
  antineutrino parameters. A search for \nue and \nuebar appearance
  has enabled a measurement of the mixing angle $\theta_{13}$. A
  measurement of the neutral-current interaction rate has confirmed
  oscillation between three active neutrino flavours. MINOS will
  continue as MINOS+ in an upgraded beam with higher energy and
  intensity, allowing precision tests of the three-flavour neutrino
  oscillation picture, in particular a very sensitive search for the
  existence of sterile neutrinos. }

\begin{document} 
\maketitle
\flushbottom

\section{Introduction}

The MINOS experiment, as an idea, was conceived in the late
1990s~\cite{ref:StanNeutrino1998}. This was a very important period in
neutrino oscillation physics. For thirty years, results from
Homestake~\cite{ref:Homestake} and the gallium
experiments~\cite{ref:Sage,ref:Gallex}, through to a number of
atmospheric neutrino
detectors~\cite{ref:Nusex,ref:Kamiokande,ref:IMB,ref:Frejus,ref:MACRO,ref:Soudan2},
had shown that neutrinos behaved in an odd fashion, often showing
significant deficits from the expected flux; but none had conclusively
determined the mechanism responsible. Then, in 1998,
Super-Kamiokande~\cite{ref:SuperKFirst} proved decisively that muon
neutrinos produced in the Earth's atmosphere disappeared as they
traveled. Around three years later, the SNO experiment showed
conclusively that neutrinos, as they propagated, changed between their
three flavours~\cite{ref:SNO}. This discovery of neutrino flavour
change showed that neutrinos had mass and did not conserve lepton
number; it was the first, and still the only, observation of physics
beyond the Standard Model.

It was during this period of discovery that the MINOS experiment was
proposed, to begin an era of precision measurement of this new
phenomenon. The data at the time were well modeled by the theory of
neutrino oscillation, in which the rate of oscillation between the
three flavours is governed by the differences between the squared
neutrino masses, $\Delta m^{2}_{21}$ and $\Delta m^{2}_{32}$. The
magnitude of the flavour change is governed by three mixing angles,
$\theta_{12}$, $\theta_{23}$ and $\theta_{13}$, and a $\mathcal{CP}$-violating
phase $\delta$; these parameters form the PMNS rotation matrix~\cite{ref:PMNS}
that relates the neutrino mass eigenstates to the flavour
eigenstates. Nature has decreed that the two mass splittings differ by
more than an order of magnitude, and that one of the mixing angles,
$\theta_{13}$, is small. Therefore, oscillation phenomenology can be
divided into two distinct regimes: `solar' oscillation driven by
$\Delta m^{2}_{21}$ and $\theta_{12}$, and `atmospheric' oscillation
driven by $\Delta m^{2}_{32}$ and $\theta_{23}$. MINOS was designed to
make precision measurements of the parameters governing the
atmospheric oscillation regime; however, it has also played an
important role in the measurement of $\theta_{13}$ and will, in the
future, make sensitive searches for the existence of sterile
neutrinos. An important feature of the MINOS design is the ability of
the detectors to identify both \numu and \numubar interactions
separately. This has allowed MINOS to make the first direct precision
tests that neutrinos and antineutrinos obey the same
oscillation parameters in the atmospheric regime~\cite{ref:MINOSRHCFirst,ref:MINOSFHCNuBars,ref:MINOSRHCSecond,ref:MINOSCC2012}.

To achieve its goals, the MINOS experiment uses the world's most
powerful neutrino beam, the NuMI beam.  In making best use of this
beam, the experiment has pioneered the two-detector technique, which
is now the gold standard for all neutrino oscillation experiments.

\section{The MINOS experiment}

The NuMI facility~\cite{ref:NuMI} provides MINOS with an intense beam
of muon flavoured neutrinos at energies of a few GeV.  The atmospheric
neutrino mass splitting drives oscillation between muon and tau
flavour neutrinos, with an energy dependence given by
\begin{equation}
P(\numu\rightarrow\numu) = 1 - \sin^{2}(2\theta)\sin^{2}
\left(
\frac{1.27\Delta m^{2}[\textrm{eV}^{2}] L[\textrm{km}]}
{E[\textrm{GeV}]}
\right).\label{eqn:NuMuDisappearance}
\end{equation}
In this two-flavour approximation, $\Delta m^{2}$ is an admixture of
$\Delta m^{2}_{32}$ and $\Delta m^{2}_{31}$; $\theta$ is also an
admixture of the mixing angles, but is heavily dominated by
$\theta_{23}$. Since MINOS cannot observe the \nutau appearance, it is
the measurement of this \numu survival probability that is used to
determine the parameters $\theta$ and $\Delta
m^{2}$~\cite{ref:MINOSCC2006,ref:MINOSCCPRD,ref:MINOSCC2008,ref:MINOSCC2010,ref:MINOSCC2012}.

A non-zero $\theta_{13}$ causes a small amount of \nue appearance in
the beam, with an energy dependence given by
\begin{equation}
P(\numu\rightarrow\nue) \approx \sin^{2}(\theta_{23})\sin^{2}(2\theta_{13})
\sin^{2}
\left(
\frac{1.27\Delta m^{2}[\textrm{eV}^{2}] L[\textrm{km}]}
{E[\textrm{GeV}]}
\right).\label{eqn:NuEAppearance}
\end{equation}
MINOS has selected a sample of \nue-enhanced events to make a
measurement of $\theta_{13}$~\cite{ref:MINOSNuEFirst,ref:MINOSNuEPRD,ref:MINOSNuE2011,ref:MINOSNuE2012}.

An important signature of neutrino oscillation is that the rate of
neutral-current (NC) neutrino interactions is unchanged by the
process.  The NC interaction is equally sensitive to all three
neutrino flavours, so this proves that flavour change is occurring
between the three active neutrino flavours.  By analysing NC
interactions, MINOS has confirmed that oscillation is the correct
picture, and has shown no evidence that this oscillation includes
additional, sterile neutrino flavours~\cite{ref:MINOSNCFirst,ref:MINOSNCPRD,ref:MINOSNC2011}.

\begin{figure}
\centering
\includegraphics[width=\textwidth]{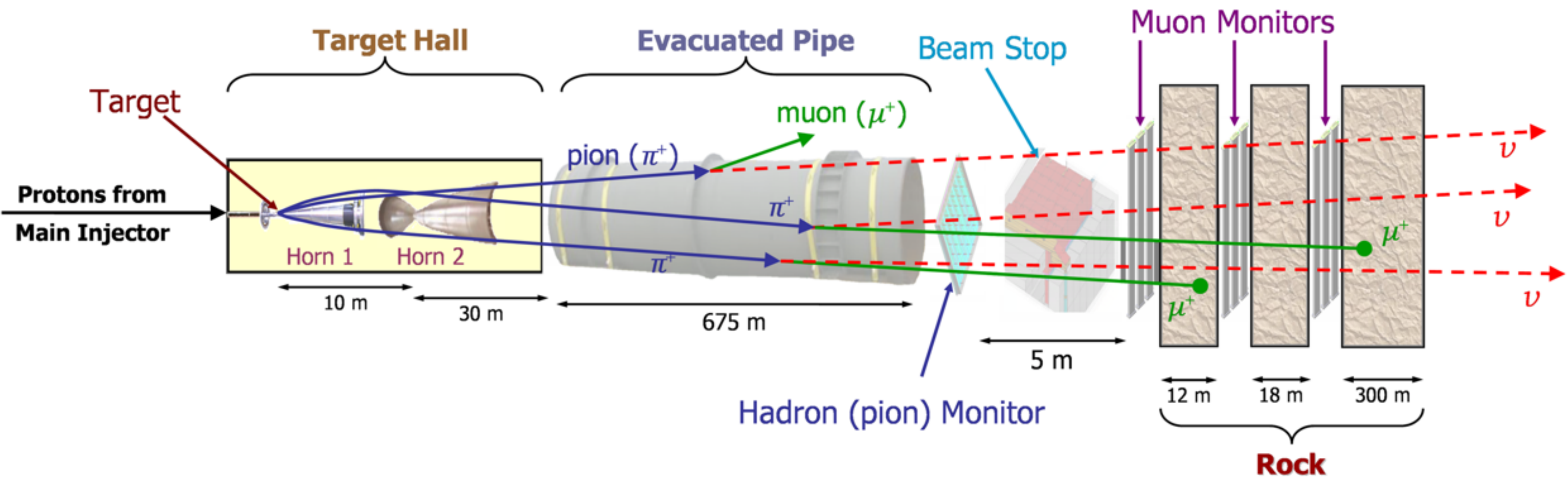}
\caption{The NuMI beam.}
\label{fig:NuMIBeam}
\end{figure}

The NuMI beam~\cite{ref:NuMI}, based at Fermilab in Chicago, has run
since 2005 and has reached a typical beam power of \unit[350]{kW}.
The Fermilab Main Injector produces a $\unit[10]{\mu s}$ pulse of
around $3\times10^{13}$ protons every \unit[2.2]{s}. These protons
have an energy of \unit[120]{GeV} and strike a graphite target, as
shown in Figure~\ref{fig:NuMIBeam}.  This target has a length of 2.0
nuclear interaction lengths and consists of a series of forty-seven
\unit[2]{cm} long graphite fins, separated by \unit[0.3]{mm}.
A shower of hadrons is produced at the target, consisting primarily of
pions with a significant kaon component at higher energies.  These
hadrons pass through two parabolic, magnetic horns which focus either
positive or negative hadrons depending on the direction of the
electric current through the horns.  The focused hadrons pass down a
\unit[675]{m} long, helium filled pipe, in which they decay to produce
a beam of predominantly muon flavoured neutrinos, with a small
electron neutrino component from the decays of muons and kaons.

\begin{figure}
\centering
\includegraphics[width=0.49\textwidth]{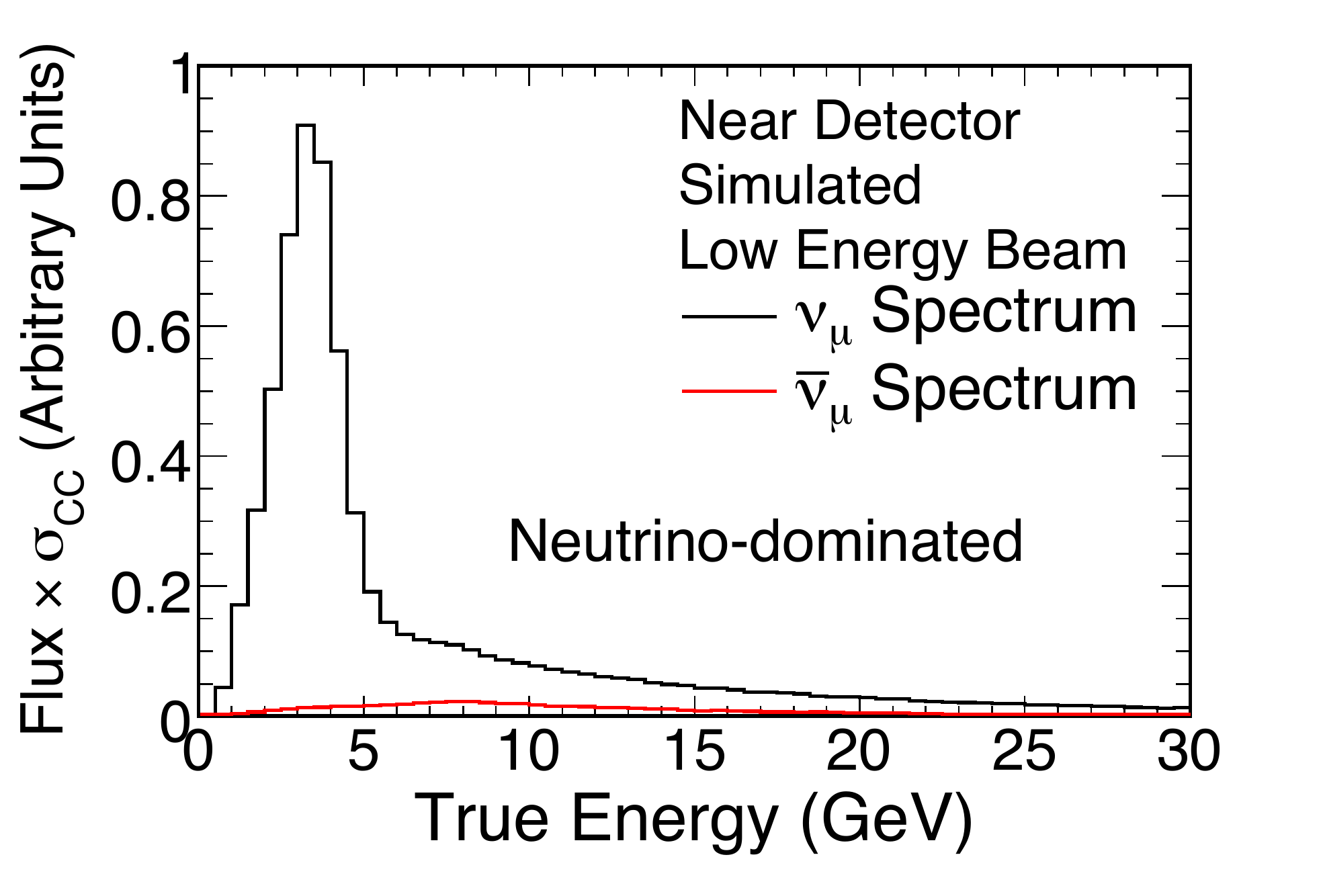}
\includegraphics[width=0.49\textwidth]{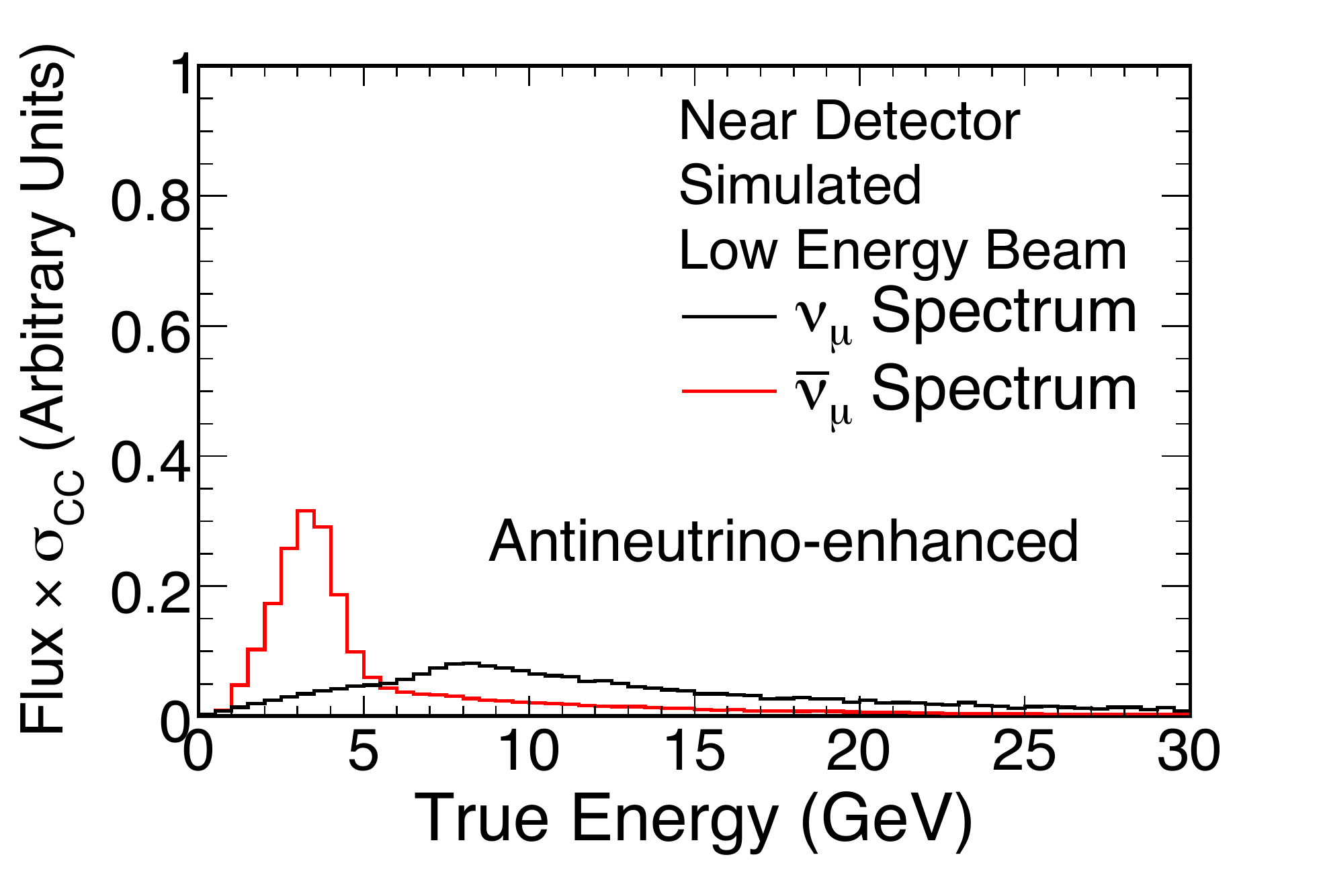}
\caption{The composition of the NuMI beam, when configured to produce (left) a neutrino-dominated beam and (right) an antineutrino-enhanced beam.}
\label{fig:BeamComposition}
\end{figure}
Figure~\ref{fig:BeamComposition} shows the composition of the NuMI
beam. With the horns configured to focus positive hadrons, the
observed beam consists of 91.7\% \numu, 7.0\% \numubar and 1.3\% \nue
and \nuebar. With the horns focusing negative hadrons, the observed
beam consists of 39.9\% \numubar, 58.1\% \numu and 2.0\% \nue and
\nuebar. The significant difference in composition and event rate
between these two configurations arises mainly from the fact that the
\numubar interaction cross section is a factor of between two and
three lower than the \numu interaction cross section.

The neutrino beam peaks at an energy of close to
\unit[3]{GeV}. However, the current through the focusing horns and the
relative positions of the horns and target are variable, allowing the
energy of the beam peak to be varied to as high as
\unit[10]{GeV}. This feature has enabled MINOS to study and understand
the beam in detail~\cite{ref:ZarkoThesis}, improving the simulation of the beam beyond the
raw Fluka~\cite{ref:Fluka} and GEANT~\cite{ref:GEANT4,ref:Flugg} Monte
Carlos, and significantly reducing the
systematic uncertainty from the modeling of the neutrino flux.

A total of $10.56\times10^{20}$ protons on target of beam data has
been analysed in the neutrino-dominated beam mode; an additional
$0.15\times10^{20}$ protons on target of data with a \unit[10]{GeV}
beam peak has also been used. In the antineutrino-enhanced beam mode,
a total of $3.36\times10^{20}$ protons on target of beam data has been
analysed.

\begin{figure}
\centering
\includegraphics[height=0.36\textwidth]{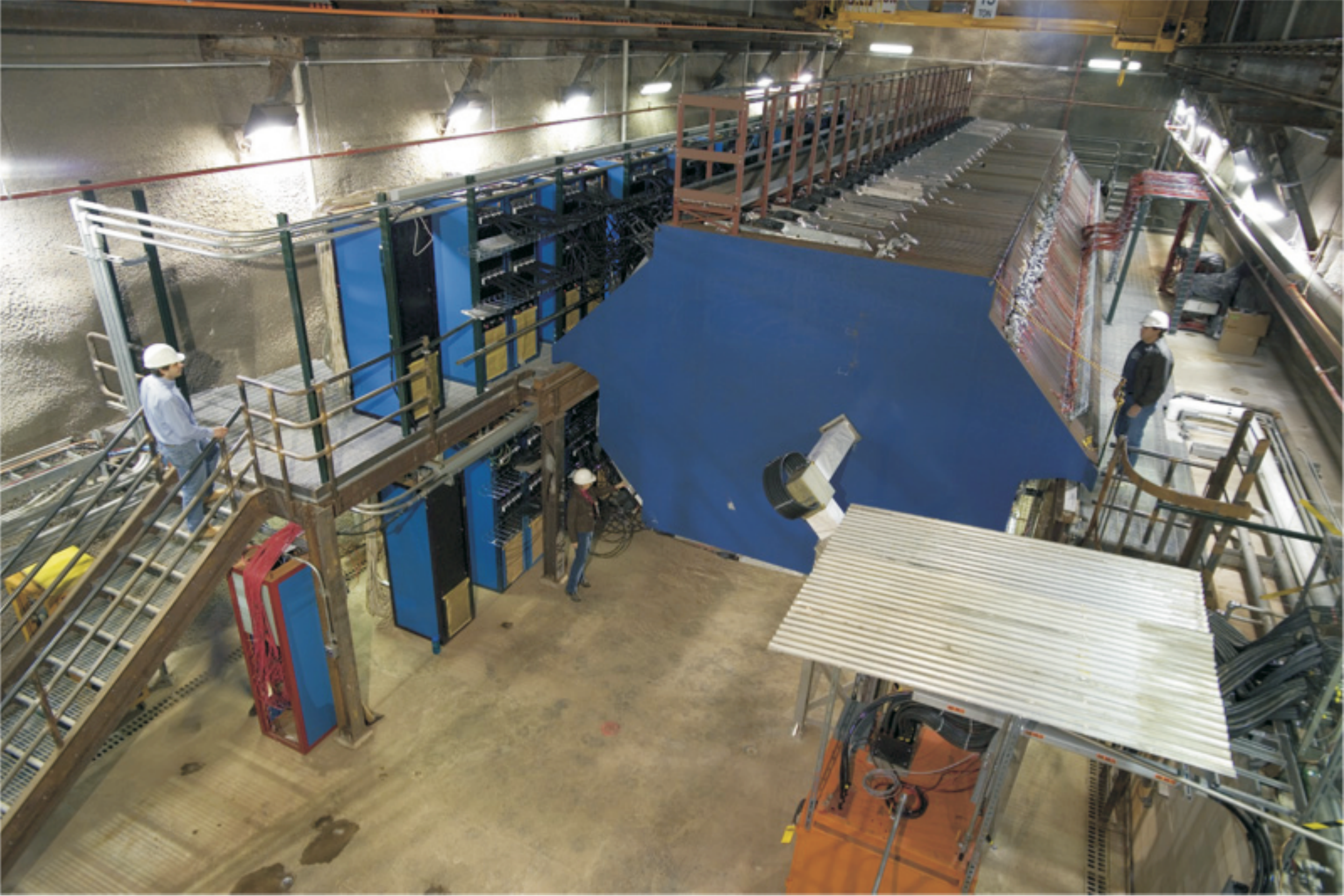}
\includegraphics[height=0.36\textwidth]{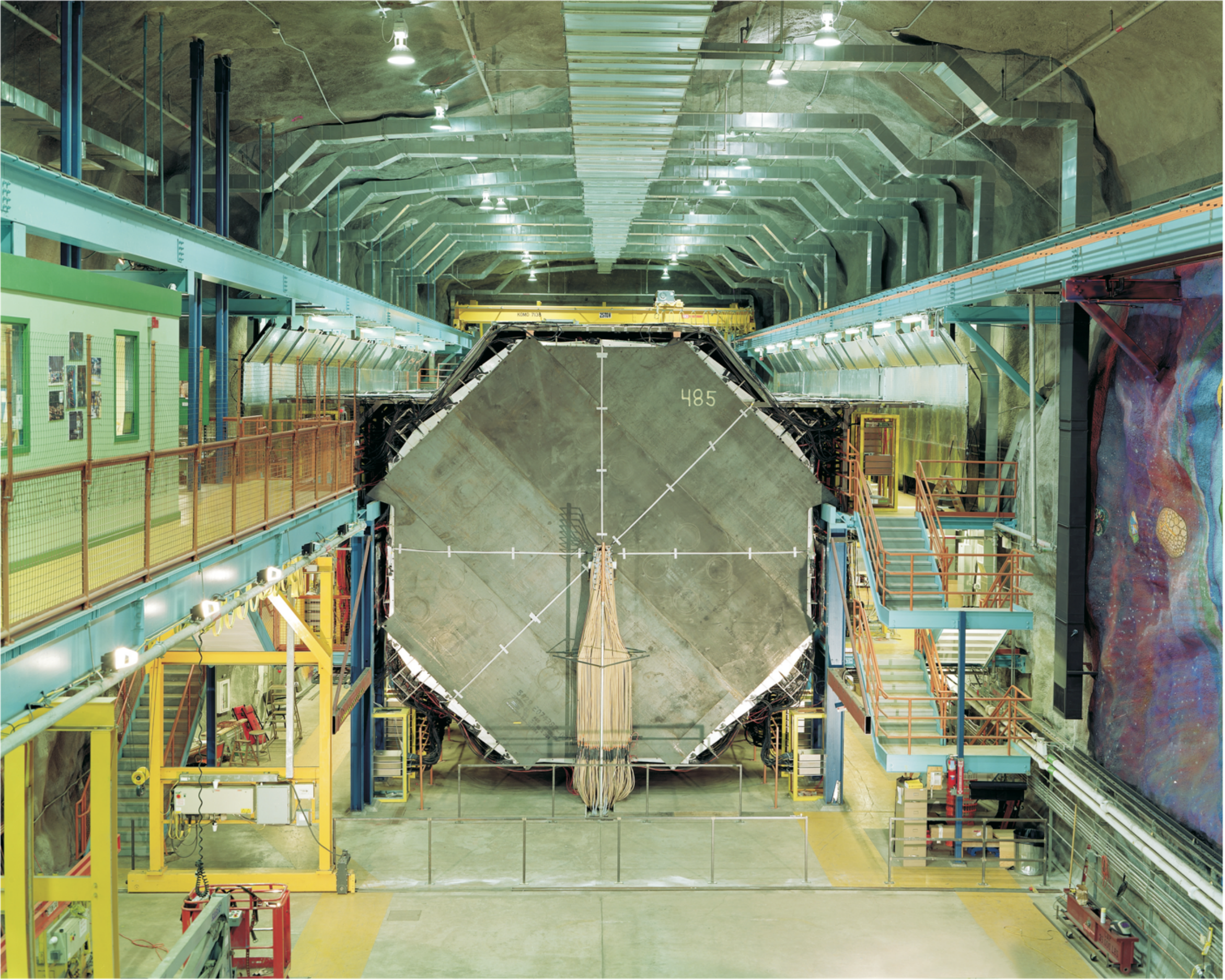}
\caption{The MINOS detectors. Left: the Near Detector at Fermilab; right: the Far Detector at the Soudan Underground Laboratory.}
\label{fig:MINOSDetectors}
\end{figure}
\begin{figure}
\centering
\includegraphics[width=0.8\textwidth]{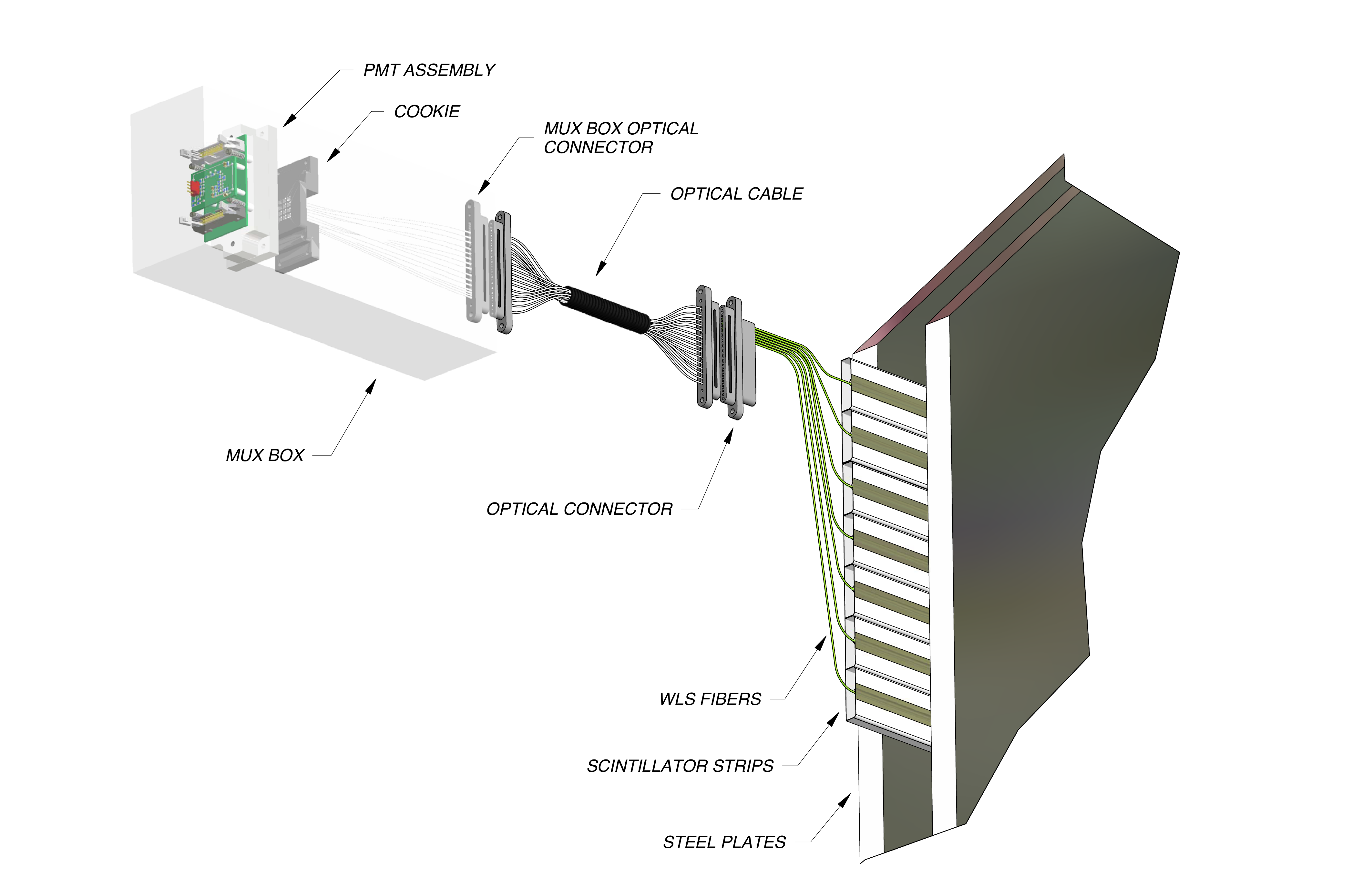}
\caption{A MINOS detector plane.}
\label{fig:MINOSPlanes}
\end{figure}

The two MINOS detectors~\cite{ref:MINOSNIM} are steel-scintillator
calorimeters, shown in Figure~\ref{fig:MINOSDetectors}.  They consist
of planes of inch-thick steel, interleaved with planes of \unit[1]{cm}
thick plastic scintillator.  The scintillator planes are divided into
\unit[4]{cm} wide strips, as shown in Figure~\ref{fig:MINOSPlanes}.
Along the centre of each strip, a wavelength shifting fibre collects
the scintillation light, shifts it to green wavelengths, and takes it
out to a photomultiplier tube.  Any charged particles passing through
the detector deposit their energy to produce light; the pattern of
these deposits allows the topology of the neutrino interaction to be
reconstructed.  The scintillator strips are aligned orthogonally on
adjacent detector planes, to allow three dimensional
reconstruction.  The detectors are magnetised to around \unit[1.3]{T},
allowing the charge of particles to be identified.

The smaller of the two detectors, the Near Detector (ND), sits at
Fermilab, \unit[1.04]{km} from the target. With a mass of
\unit[0.98]{kton}, it measures the energy spectra of the neutrinos
before oscillation.  The Far Detector is located at the Soudan
Underground Laboratory in northern Minnesota, \unit[705]{m}
underground and \unit[735]{km} from the target.  With a mass of
\unit[5.4]{kton}, it again measures the neutrino energy spectra,
seeing the appearance and disappearance of neutrinos due to
oscillation.

This two-detector arrangement is very powerful in reducing systematic
uncertainties.  Neutrino physics is beset with uncertainty: in
particular, interaction cross sections are unknown to many tens of per
cents, and neutrino fluxes can be mismodeled by similar amounts.
However, these uncertainties affect both detectors in exactly the same
way so cancel very effectively when a ratio is taken of the Near to
Far Detector energy spectra.  As an indication of how well this works,
despite the uncertainties of tens of per cent in the simulated event
rate in the detectors, once the Near to Far Detector ratio is taken,
the normalization is known to 1.6\%.

The MINOS Far Detector is also a very effective detector of neutrinos
produced in the atmosphere.  Since it was switched on in 2003, it has
recorded \unit[37.9]{kton-years} of data, recording 2072 candidate
neutrino interactions that have been included into the analyses of the
beam data to improve the precision of the oscillation parameter
measurements~\cite{ref:MINOSAtmosFirst,ref:MINOSAtmosChargeSeparated,ref:MINOSAtmosPRD,ref:MINOSCC2012}.

\section{Neutrino interactions in the MINOS detectors}

\begin{figure}
\centering
\includegraphics[width=\textwidth]{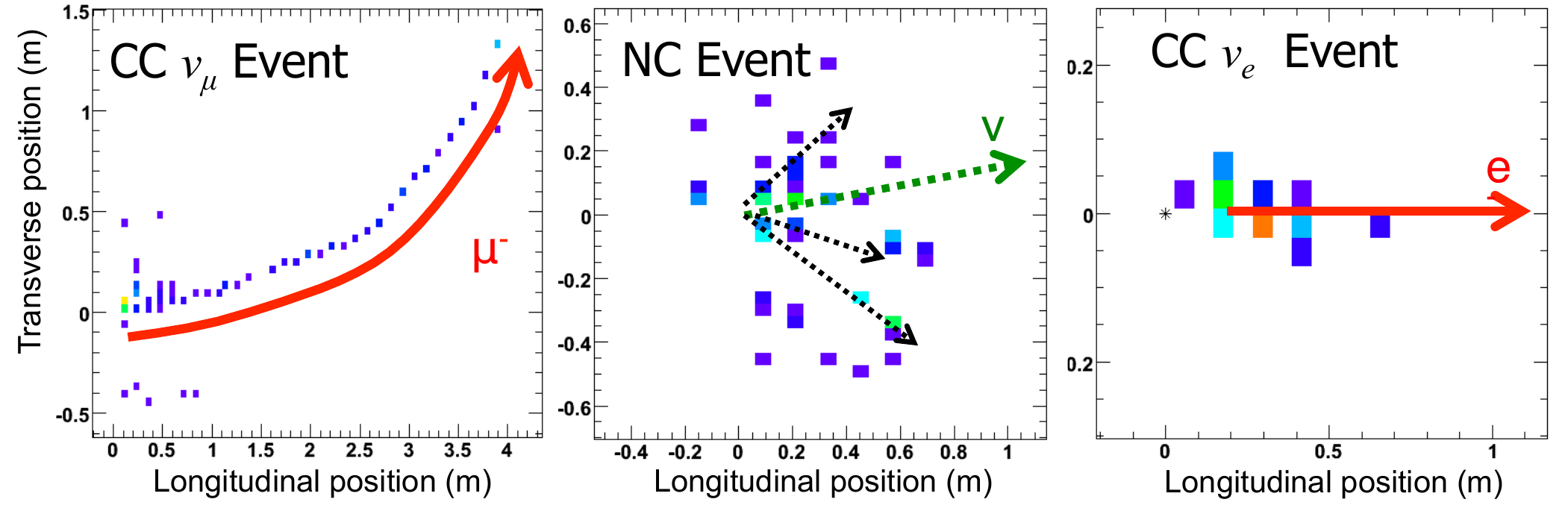}
\caption{Neutrino interaction topologies observed in the MINOS
  detectors. Left: a CC \numu interaction. Middle: a NC
  interaction. Right: a CC \nue interaction. Each coloured rectangle
  represents an excited scintillator strip, the colour indicating the
  amount of light: purple and blue are low light levels, through to
  orange and red for the highest light levels.}
\label{fig:EventDisplays}
\end{figure}

Three types of neutrino interaction, shown in Figure~\ref{fig:EventDisplays}, are of interest to MINOS.
Muon neutrinos and antineutrinos interact through the charged-current (CC) process
\[
\numu(\numubar)+X\rightarrow\mu^{-(+)}+X^{\prime}.
\]
The cascade of hadrons, $X^{\prime}$, produces a diffuse shower of energy deposits near the interaction vertex.
The muon produces a long track that curves in the magnetic field, the direction of curvature identifying the incoming neutrino as a \numu or a \numubar.

All active neutrino flavours undergo NC interactions through the process
\[
\nu+X\rightarrow\nu+X^{\prime}.
\]
Only the hadronic cascade is observed, producing a diffuse shower of energy deposits.

Finally, electron neutrinos undergo CC interactions through the process
\[
\nu_{e}+X\rightarrow e^{-}+X^{\prime}.
\]
The electron gives rise to an electromagnetic shower, which produces a much denser, more compact shower of energy deposits.

The energy of the neutrino is determined by summing the energies of
the shower and any muon track. The muon energy is determined from the
length of stopping tracks, leading to a resolution of around 5\%, and
from the curvature in the magnetic field for tracks that exit the
detector, leading to a resolution of around 10\%. For NC and \nue CC
interactions, the energy of the shower is determined through
calorimetry. The calorimetric energy resolution for hadronic showers
is around $55\%/\sqrt{\textrm{energy}}$~\cite{ref:MikeThesis} and for
electromagnetic showers
$20\%/\sqrt{\textrm{energy}}$~\cite{ref:TriciaThesis}. For \numu CC
interactions, a more sophisticated approach is used to improve the
resolution of hadronic shower energy
measurement~\cite{ref:BackhouseThesis}. For low energy showers (of a
few GeV or below), significant additional information is held in the
topology of the shower. Three event characteristics are used: the
calorimetric energy deposit within \unit[1]{m} of the interaction
vertex, the sum of the calorimetric energy in the two largest showers
in the event, and the physical length of the largest shower. These
variables are input into a $k$-nearest-neighbour
algorithm~\cite{ref:kNN}, which finds the best matches from a library
of simulated events and uses these to estimate the hadronic
energy. This improves the shower energy resolution from 55\% to 43\%
for showers between \unit[1.0]{GeV} and \unit[1.5]{GeV}.

\subsection{Charged-current \numu and \numubar interactions}

\begin{figure}
\centering
\includegraphics[height=0.315\textwidth]{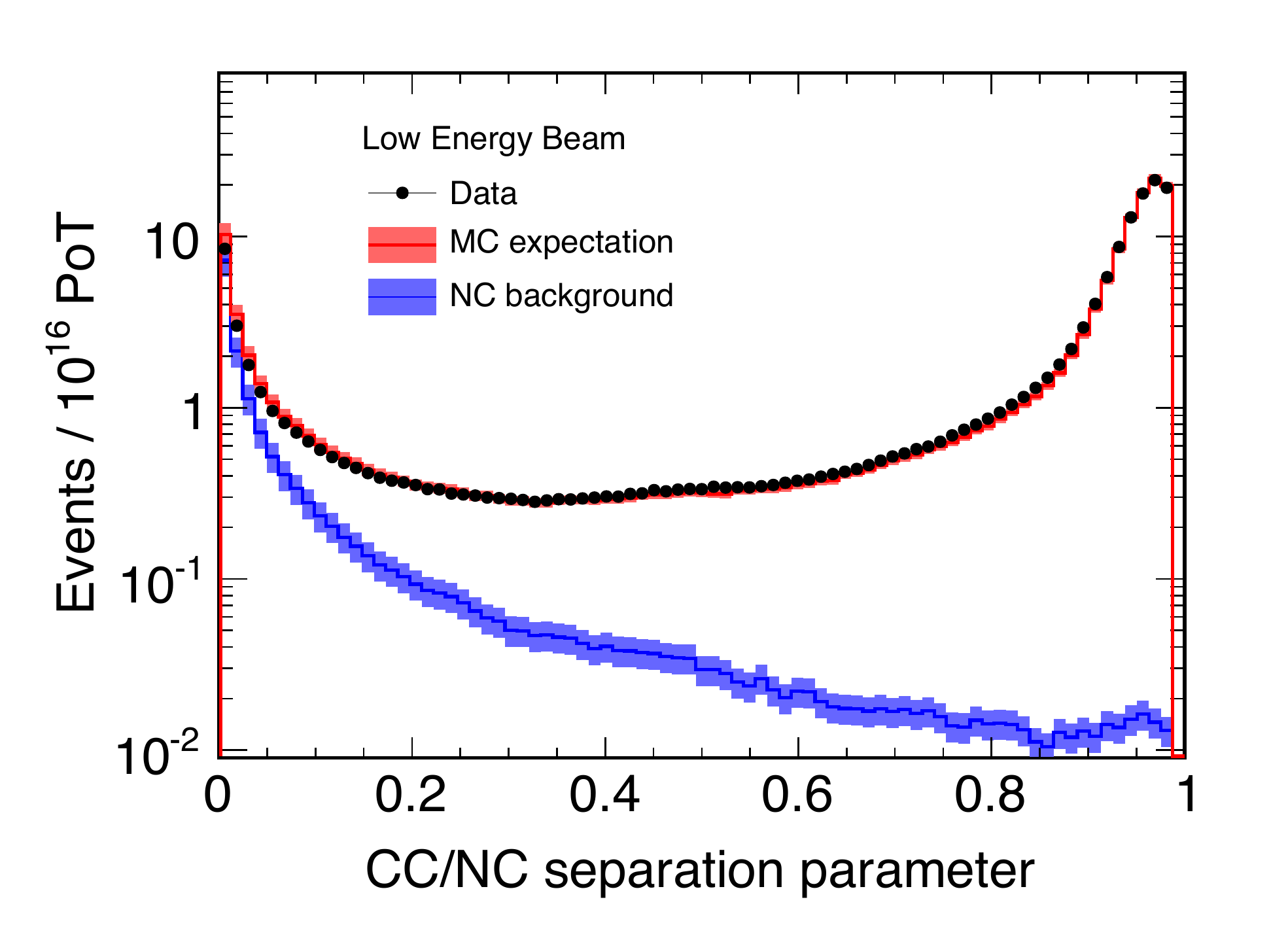}
\includegraphics[height=0.365\textwidth]{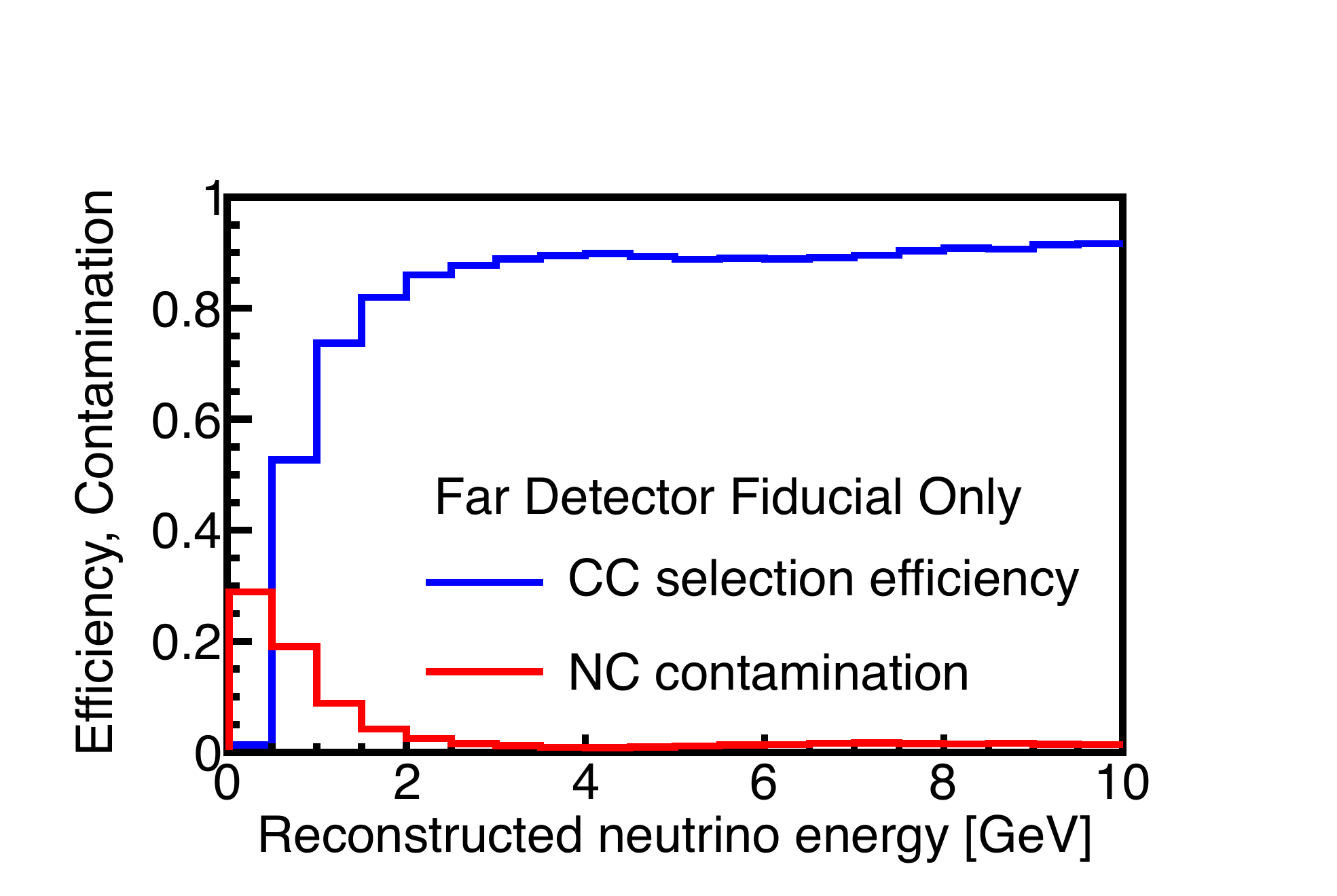}
\caption{Left: the discrimination variable used to separate \numu CC interactions from hadronic backgrounds. Events with a parameter value greater than 0.3 are selected as \numu CC interactions. Right: the efficiency and background contamination of the selected \numu CC sample in the Far Detector.}
\label{fig:roID}
\end{figure}

To make a measurement of $P(\numu\rightarrow\numu)$, it is necessary
to select a pure sample of \numu CC interactions. This is achieved by
selecting events with a clear muon track. The main loss in efficiency
comes from events with a high inelasticity in which a short muon track
is hidden in a large hadronic cascade. The main background occurs at
low energies, and consists of small cascades from NC interactions in
which a low energy hadron, such as a proton or a charged pion,
exhibits a track-like topology that mimics a low energy muon. Four
variables are constructed that discriminate between muons tracks,
which are typically long and show a constant energy deposition along
the length, and spurious hadronic tracks, which are typically shorter
and show greater fluctuations in the energy deposition. These
variables are the event length, the average energy deposited per
scintillator plane along the track, the transverse energy deposition
profile, and the fluctuation of the energy deposition along the
track. These variables are input into a $k$-nearest-neighbour
algorithm, which calculates a single discrimination variable, shown in
Figure~\ref{fig:roID}~\cite{ref:RustemThesis}. Events for which this
variable is greater than 0.3 are selected as CC \numu interactions,
yielding a sample with 90\% efficiency; below \unit[2]{GeV}, the NC
contamination is 6.5\%. The energy dependent efficiency and
contamination are shown in Figure~\ref{fig:roID}.

The CC interactions of \numu and \numubar result in very similar
topologies; the $k$-nearest-neighbour discriminant is therefore used
in the same way in both the neutrino-dominated and
antineutrino-enhanced beams. When performing a direct measurement of
the antineutrino oscillation parameters, an additional selection cut
is made, requiring the charge of the muon track to be positive. This
uses the direction of curvature of the muon as measured by a Kalman
Filter algorithm~\cite{ref:JohnMarshallThesis}. A further sample of
\numubar CC interactions is obtained from the 7\% \numubar component
in the neutrino-dominated beam. This sample contains a significant
background of \numu events in which a $\mu^{-}$ has been identified
with the incorrect charge, often at low energies where the muon
undergoes significant scattering. Therefore a much stricter set of
selection criteria are applied to purify this \numubar
sample~\cite{ref:MINOSFHCNuBars}.

\subsection{Charged-current \nue interactions}

\begin{figure}
\centering
\includegraphics[width=0.49\textwidth]{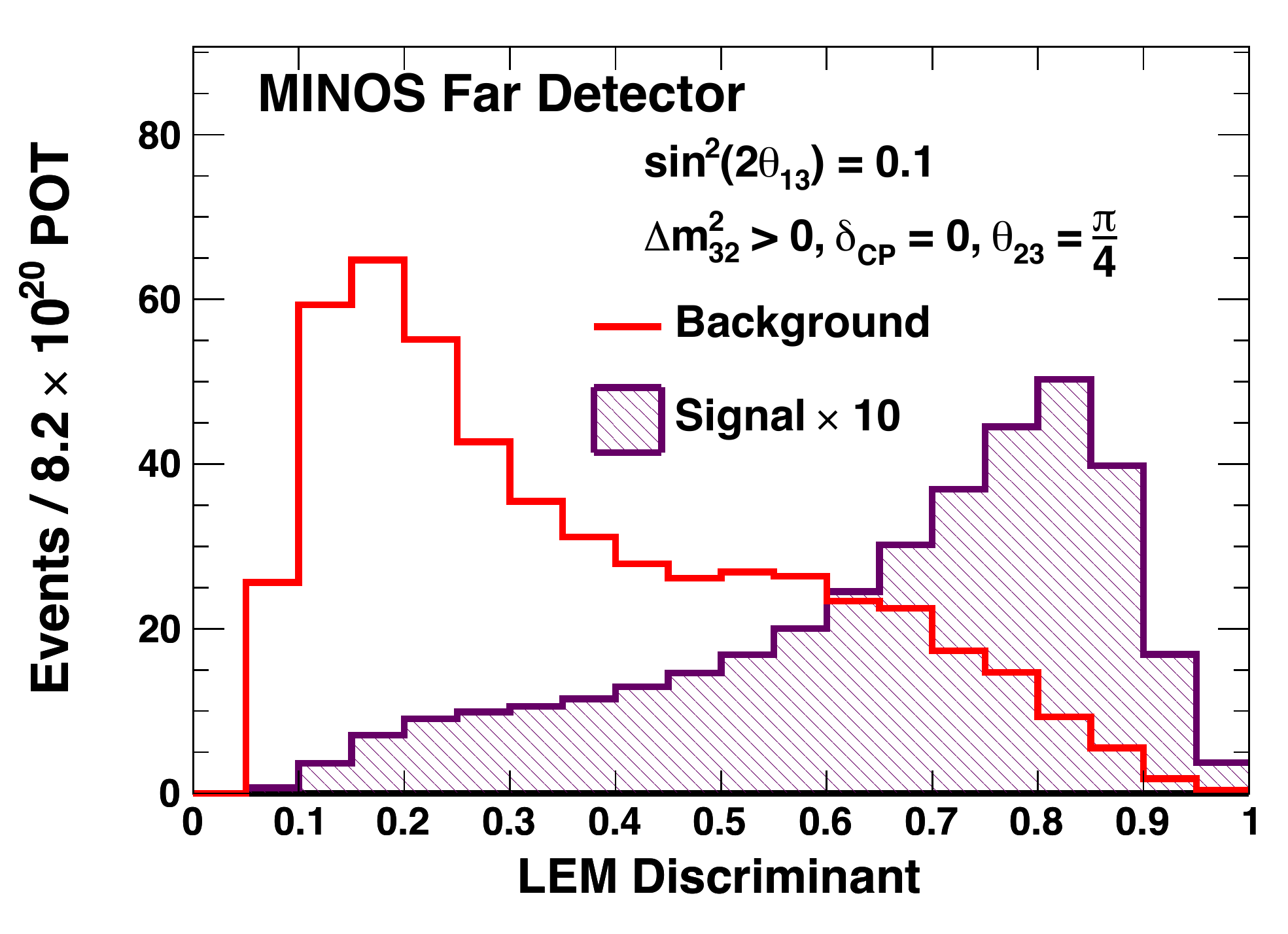}
\includegraphics[width=0.49\textwidth]{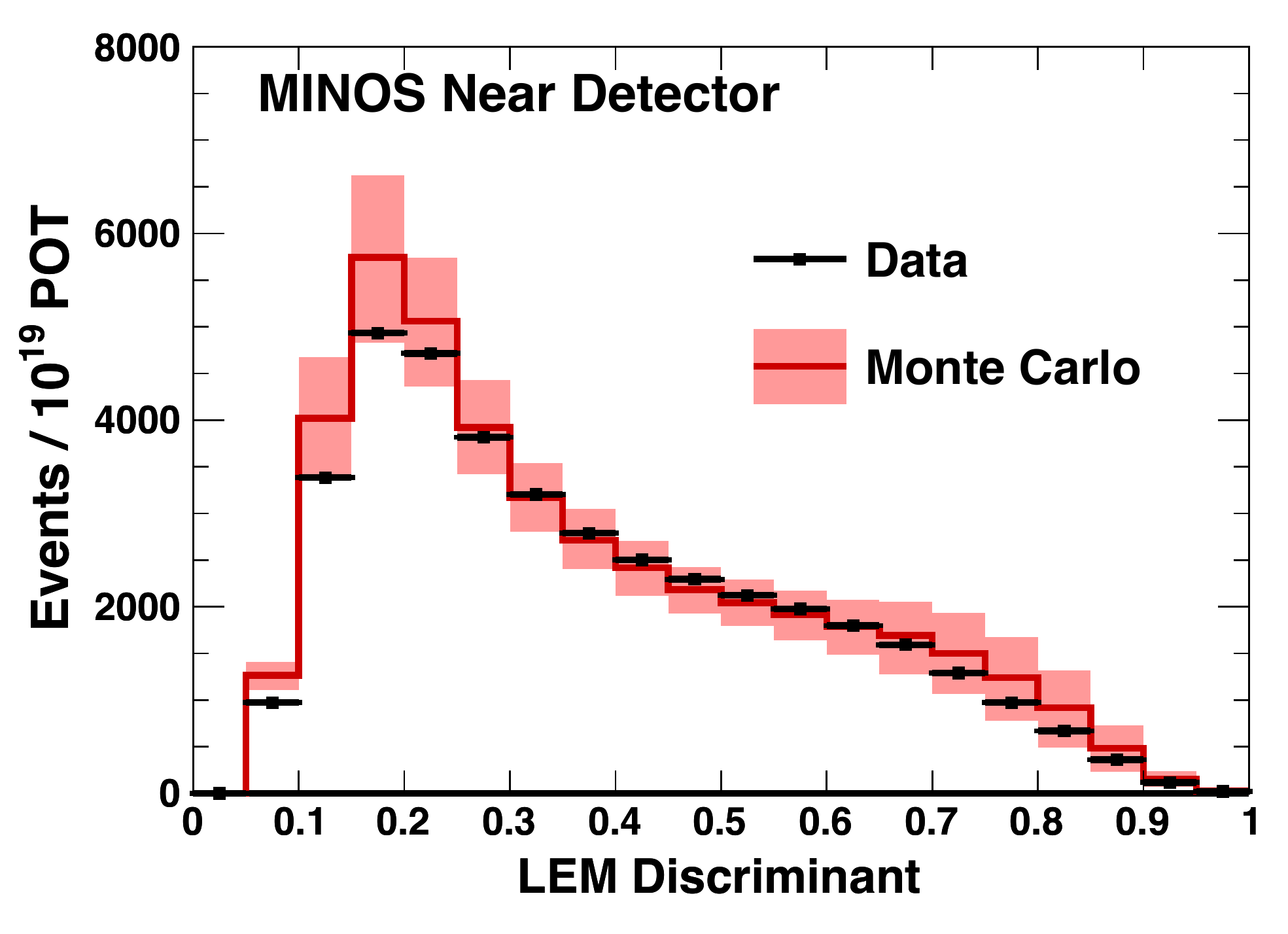}
\caption{Left: the Library Event Matching discriminant, showing the
  expected distribution for background and CC \nue signal events in
  the Far Detector, in the neutrino-dominated beam. Note that the
  signal, simulated for $\sin^{2}(2\theta_{13})=0$, $\delta=0$ and a
  normal mass hierarchy, has been scaled up by a factor of ten for
  visibility. Right: the same discriminant as observed in the Near
  Detector, compared with the simulated expectation.}
\label{fig:NuESelector}
\end{figure}

The selection of \nue CC interactions focuses on identifying the dense
showers from the electromagnetic interaction of the electron, rather
than the much more diffuse hadronic showers. The primary background
comes from purely hadronic showers which can have a denser than
average energy deposit, particularly in the presence of a neutral pion
decaying to photons. Once a set of shower-like events in the signal
region of \unit[1--8]{GeV} has been obtained, a pattern matching
approach, called Library Event Matching, is used to identify the interactions most
likely to be \nue CC~\cite{ref:PedroThesis,ref:RuthThesis}. Each
event in the data is compared to a library of $5\times10^{7}$
simulated signal and background events; its similarity to the library
events is quantified by comparing the pattern of energy deposits in
each scintillator strip excited by the shower, where the energy
deposit is quantified by the charge recorded on the photomultiplier
tube. For an arbitrary energy deposit, the mean expected charge on a
photomultiplier tube will be some value $\lambda$. The probability of
observing an amount of charge $n$ is then a Poisson distribution,
$P(n|\lambda)$. The likelihood, $\mathcal{L}$, of a data event
corresponding to the same physical shower topology as a simulated
library event can therefore be calculated as
\[
\log\mathcal{L} = \sum_{i=1}^{N_{\mathrm{strips}}}
\log\left[
\int^{\infty}_{0}
P(n_{\mathrm{data}}^{i}|\lambda)
P(n_{\mathrm{lib}}^{i}|\lambda)
\mathrm{d}\lambda
\right],
\]
where $i$ represents the $i$th scintillator strip in the shower. Using
this definition of the likelihood, the 50 library events are
identified that best match the data event. Three quantities are
calculated from this set of 50 best-matching library events: the
fraction that are true \nue CC events, the average inelasticity of the
true \nue CC events, and the average fraction of charge that overlaps
between the data event and each \nue CC library event. These three
quantities are input to a neural network, which calculates a
classification variable shown in Figure~\ref{fig:NuESelector}. Events
with a classification variable value above 0.6 are selected for
analysis.

The efficiency of the \nue CC selection is estimated from the data,
rather than relying totally on the simulation. To obtain a pure sample
of true hadronic showers a sample of well identified \numu CC events
is selected, and the energy depositions corresponding to the muon track
are removed~\cite{ref:AnnaThesis}. The simulated energy depositions of
an electron are then inserted~\cite{ref:JoshThesis}, providing a
realistic sample of \nue CC events. Using this method, the \nue CC
identification efficiency is found to be $(57.4\pm2.8)\%$ in the
neutrino-dominated beam, and $(63.3\pm3.1)\%$ in the
antineutrino-enhanced beam.

\subsection{Neutral-current interactions}

\begin{figure}
\centering
\includegraphics[width=0.7\textwidth]{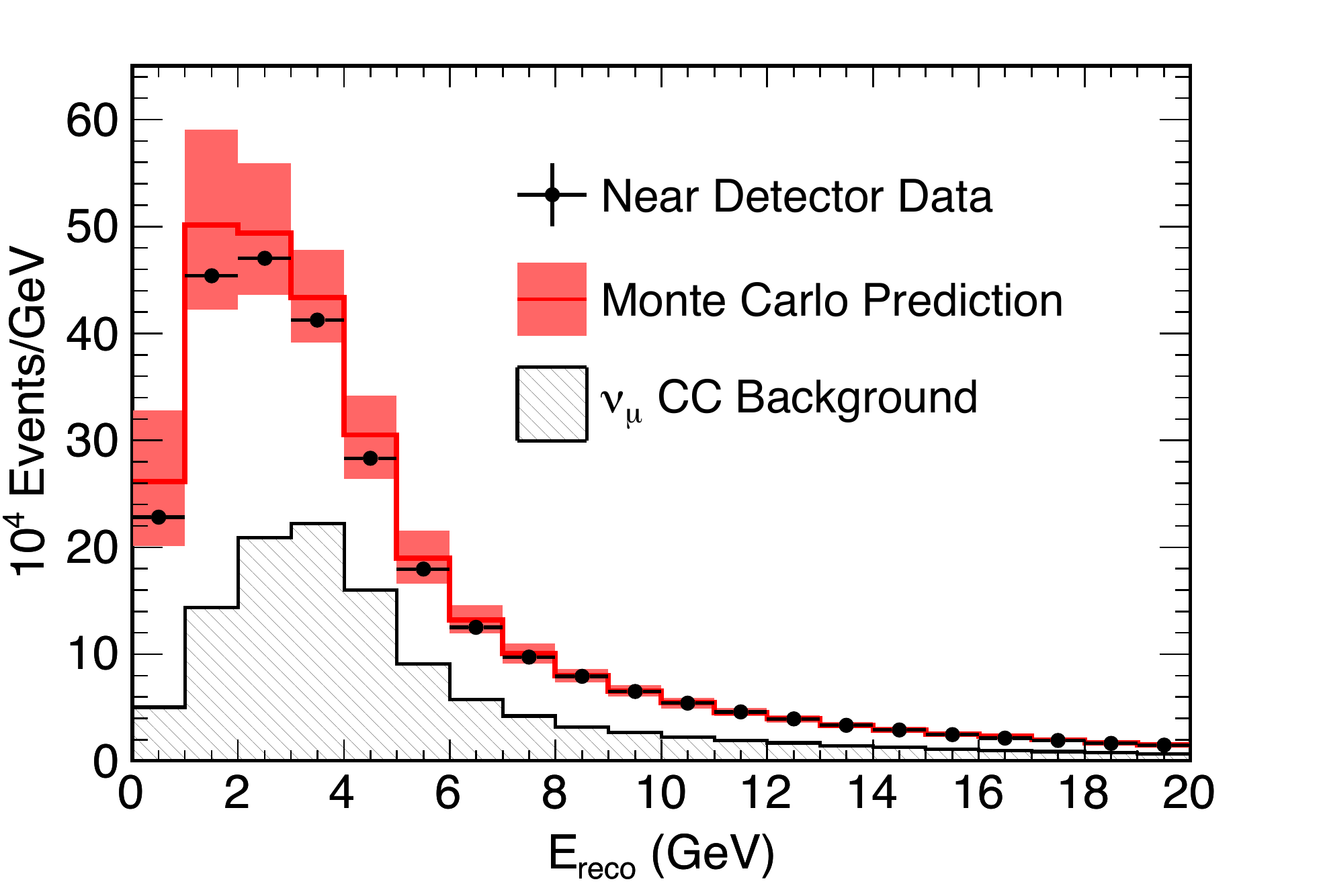}
\caption{The sample of events identified as NC interactions in the Near Detector.}
\label{fig:NCND}
\end{figure}

The signal of an NC interaction is a diffuse hadronic shower. \numu CC
interactions also produce hadronic showers, and if the inelasticity
is high, the tell-tale muon track may not visibly extend past the
shower. To purify a sample of NC interactions, a simple cut-based
approach is taken~\cite{ref:GemmaThesis}: events are classified as
NC-like if the event contains no reconstructed track, or if the track
extends no more than six planes past the end of the shower. The
resulting distribution of NC interactions in the Near Detector is
shown in Figure~\ref{fig:NCND}. The NC identification efficiency is
89\%, with 61\% purity. This selection will identify 97\% of \nue CC
interactions as NC events; therefore an analysis of NC interactions in
the FD must account for the \nue appearance caused by a non-zero
$\theta_{13}$.

\subsection{Atmospheric neutrinos}

Atmospheric neutrino interactions are selected out of any activity
seen in the FD outside of the $\unit[10]{\mu s}$ periods when the NuMI
beam is active~\cite{ref:MINOSAtmosPRD}.  The oscillation signal is
contained in the \numu CC interactions, and as with the beam-induced
interactions these are identified by the presence of a muon track.

The FD has a single-hit timing resolution of \unit[2.5]{ns}. This
timing information is used to determine the direction in which the
detector activity is traveling.  Any downwards traveling activity is
required to begin well inside the detector, to eliminate cosmic muons
entering from above.  All upwards or horizontally traveling activity
is almost certain to be neutrino-induced, since no other particle can
survive through the many kilometres of rock.  All activity with a
zenith angle of $\cos\theta_{z}<0.14$ is defined as horizontal or
down-going; this corresponds to an overburden of at least \unit[14]{km}
water equivalent.

From this sample of neutrino-induced activity, all events with a track
crossing at least eight planes are designated track-like; all events
with only shower-like activity crossing at least four planes are
designated shower-like.  These track-like and shower-like samples are
used in the neutrino oscillation measurements. The track-like sample
contains the oscillation signal of \numu disappearance. The
shower-like sample contains mainly NC interactions and \nue and
\nuebar CC interactions; it shows little oscillation signal, but is
very important for setting the normalization of the atmospheric
neutrino flux.

\section{Muon neutrino disappearance}

\begin{figure}
\centering
\includegraphics[width=0.7\textwidth]{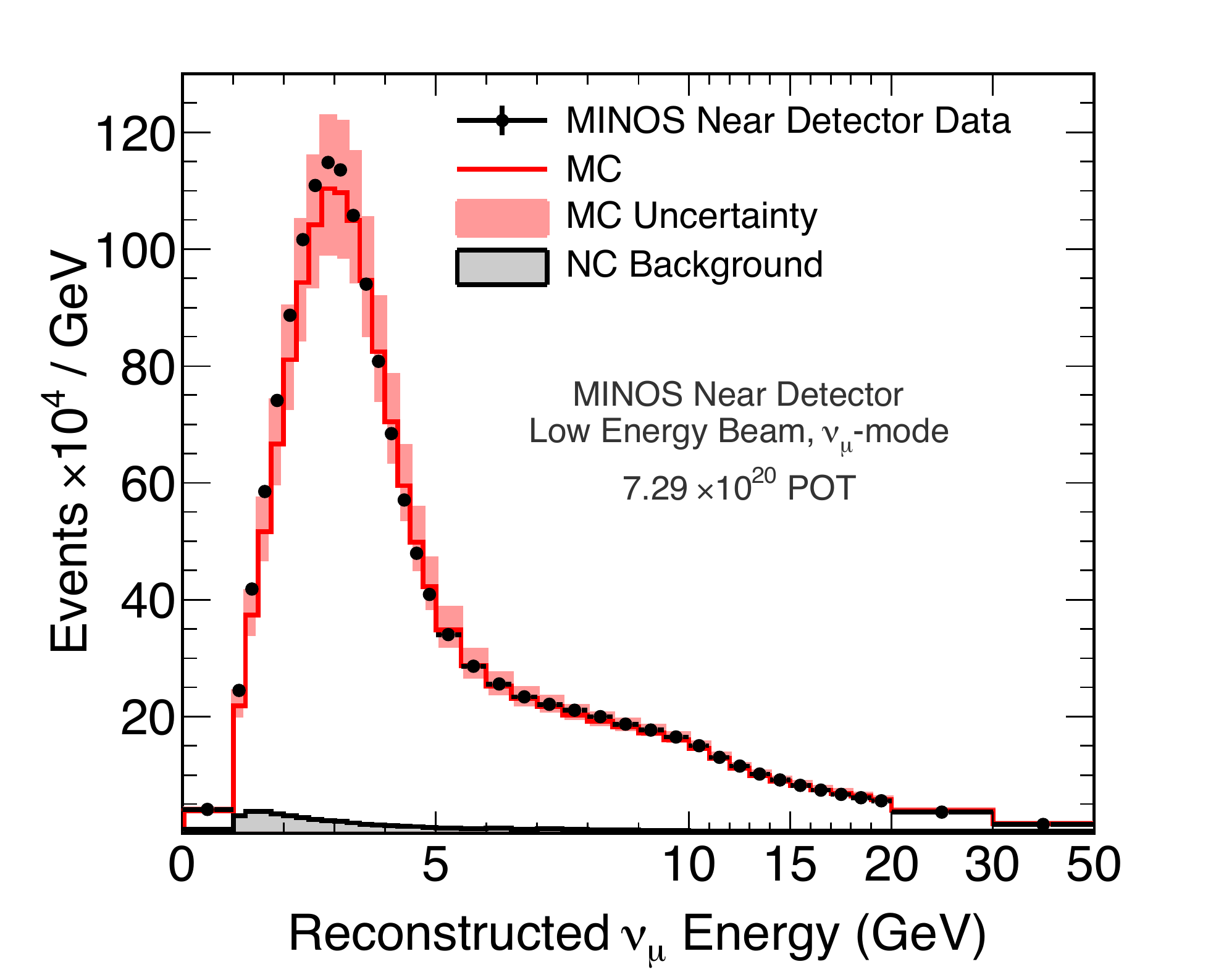}
\caption{The energy spectrum of \numu CC interactions observed in the
  ND, compared to the simulation.}
\label{fig:CCNDSpectrum}
\end{figure}

The atmospheric oscillation parameters, $|\Delta m^{2}|$ and
$\sin^{2}(2\theta)$, are measured by observing and fitting the energy
dependence of \numu and \numubar disappearance.  To minimise the
impact of systematic uncertainties, the energy spectra of the \numu
and \numubar CC interactions observed in the ND (shown in
Figure~\ref{fig:CCNDSpectrum} for the neutrino-dominated beam) are used
to predict the spectrum at the FD, in the absence of
oscillation~\cite{ref:MINOSCCPRD,ref:MyThesis}.  The neutrino energy
spectra at the ND and FD are not identical: the ND subtends a
relatively large angle to the beam, so for each pion or kaon a range
of decay angles can produce a neutrino that passes through the
detector, corresponding to a range of neutrino energies. However, the
FD is effectively a point when viewed from the neutrino production
location, so a single decay angle for each hadron, and therefore a single
neutrino energy, contributes to the flux.  To take this difference
into account, the hadron-decay kinematics are encoded into a beam
transfer matrix that converts the observed ND flux into a predicted
FD flux.  Once the ND data has been used in this way, the most
important systematic uncertainties are those that can affect the two
detectors differently, primarily reconstruction efficiencies and
miscalibrations of the neutrino energy measurement in the
detectors~\cite{ref:StephenThesis}. These uncertainties are included
in the fit that extracts the oscillation
parameters~\cite{ref:JessThesis}.

\begin{figure}
\centering
\includegraphics[width=\textwidth]{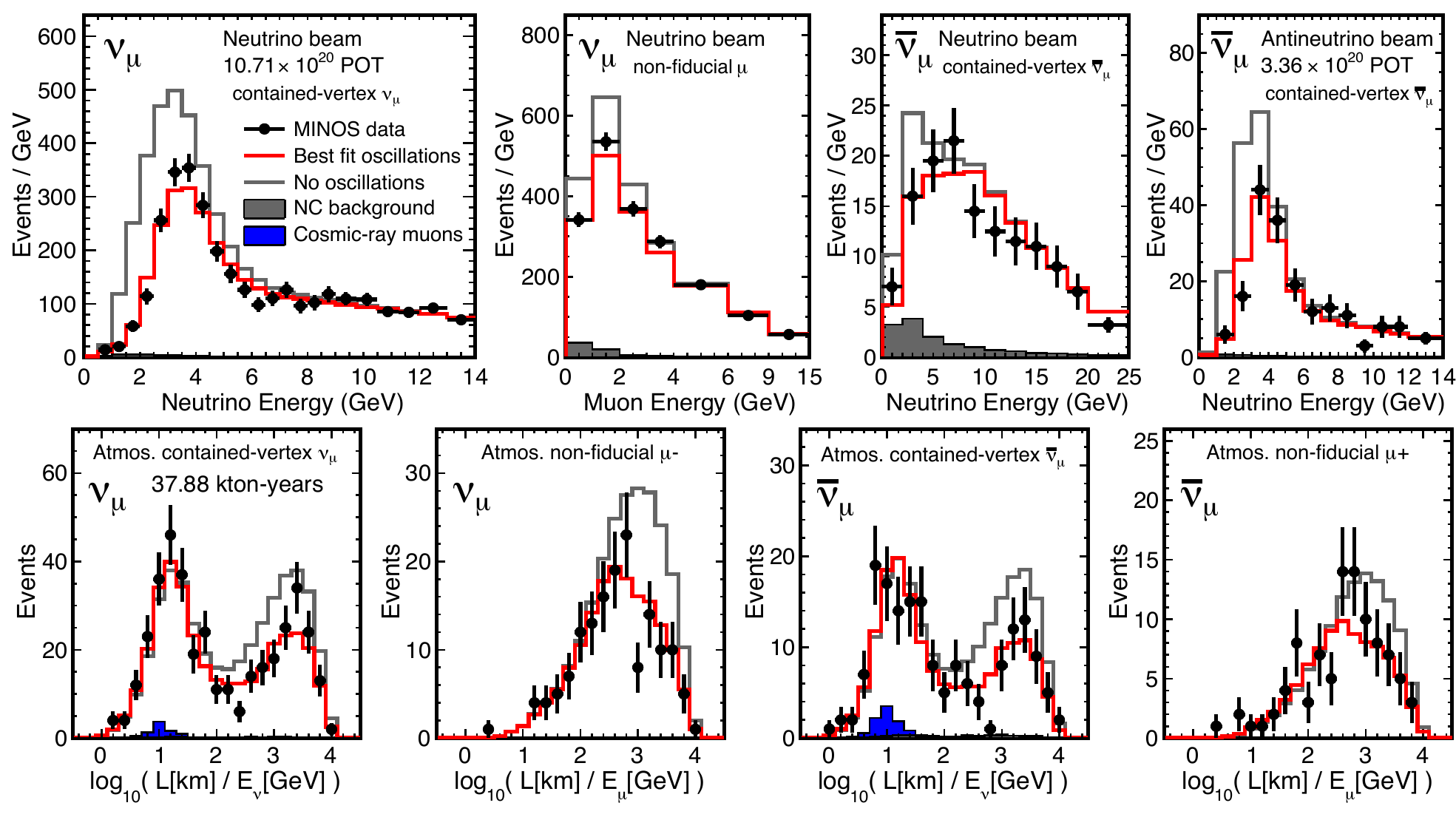}
\caption{The energy spectra of \numu and \numubar CC interactions
  observed at the FD, compared to the expectation with and without
  oscillation. The top row shows beam-induced neutrinos; the bottom
  row shows atmospheric neutrinos.}
\label{fig:CCFDSpectrum}
\end{figure}

\begin{figure}
\centering
\includegraphics[width=0.7\textwidth]{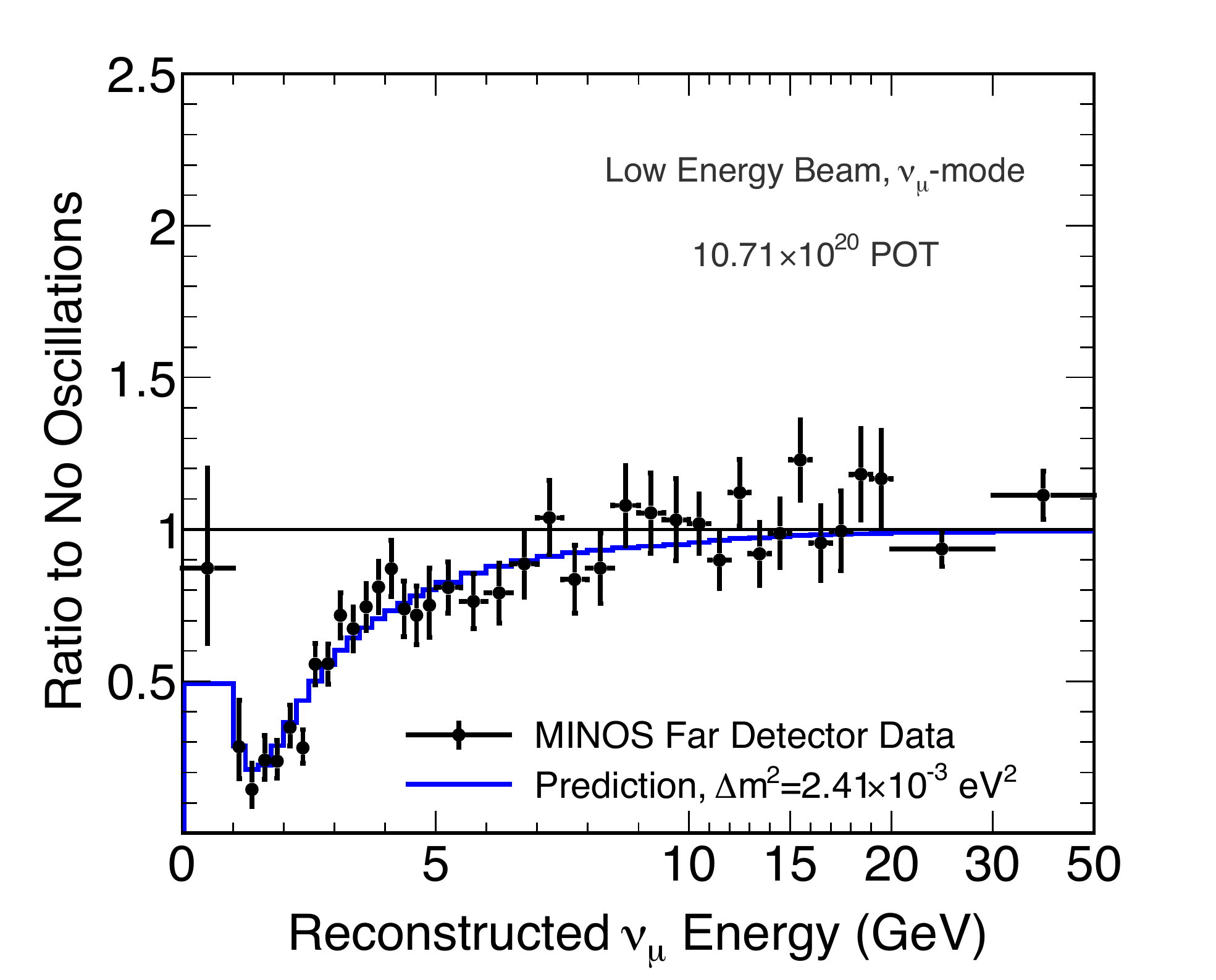}
\caption{The ratio of the observed \numu energy spectrum to the
  expectation in the case of no oscillation, in the
  neutrino-dominated beam.}
\label{fig:CCFDRatio}
\end{figure}

The top row of Figure~\ref{fig:CCFDSpectrum} shows the predicted
spectra of \numu and \numubar CC interactions from the
neutrino-dominated and antineutrino-enhanced beams at the FD, along
with the data. In the neutrino-dominated beam, an additional sample is
used, consisting of neutrinos interacting outside the fiducial volume
of the detector, and in the rock surrounding the
detector~\cite{ref:AaronThesis,ref:StraitThesis}. This non-fiducial
sample consists mainly of high energy neutrinos, and has significantly
lower resolution as not all the energy is contained in the detector;
however, it does contain some oscillation information. In all samples,
a clear, energy-dependent deficit of \numu and \numubar interactions is
observed.  The ratio of the data to the expectation for the \numu
interactions in the neutrino-dominated beam is shown in
Figure~\ref{fig:CCFDRatio}. This ratio shows the `dip and rise' energy
dependence of the deficit, which is characteristic of oscillation and
described by equation~(\ref{eqn:NuMuDisappearance}).

The bottom row of Figure~\ref{fig:CCFDSpectrum} shows the spectra of
atmospheric \numu and \numubar CC interactions, as a function of $L/E$
where $L$ is the distance traveled by the neutrino and $E$ its
energy. The atmospheric neutrino events are divided into \numu and
\numubar interactions according to the direction of curvature of the
muon, and separated into samples depending on whether or not the
interaction vertex is contained in the detector.

\begin{figure}
\centering
\includegraphics[width=0.7\textwidth]{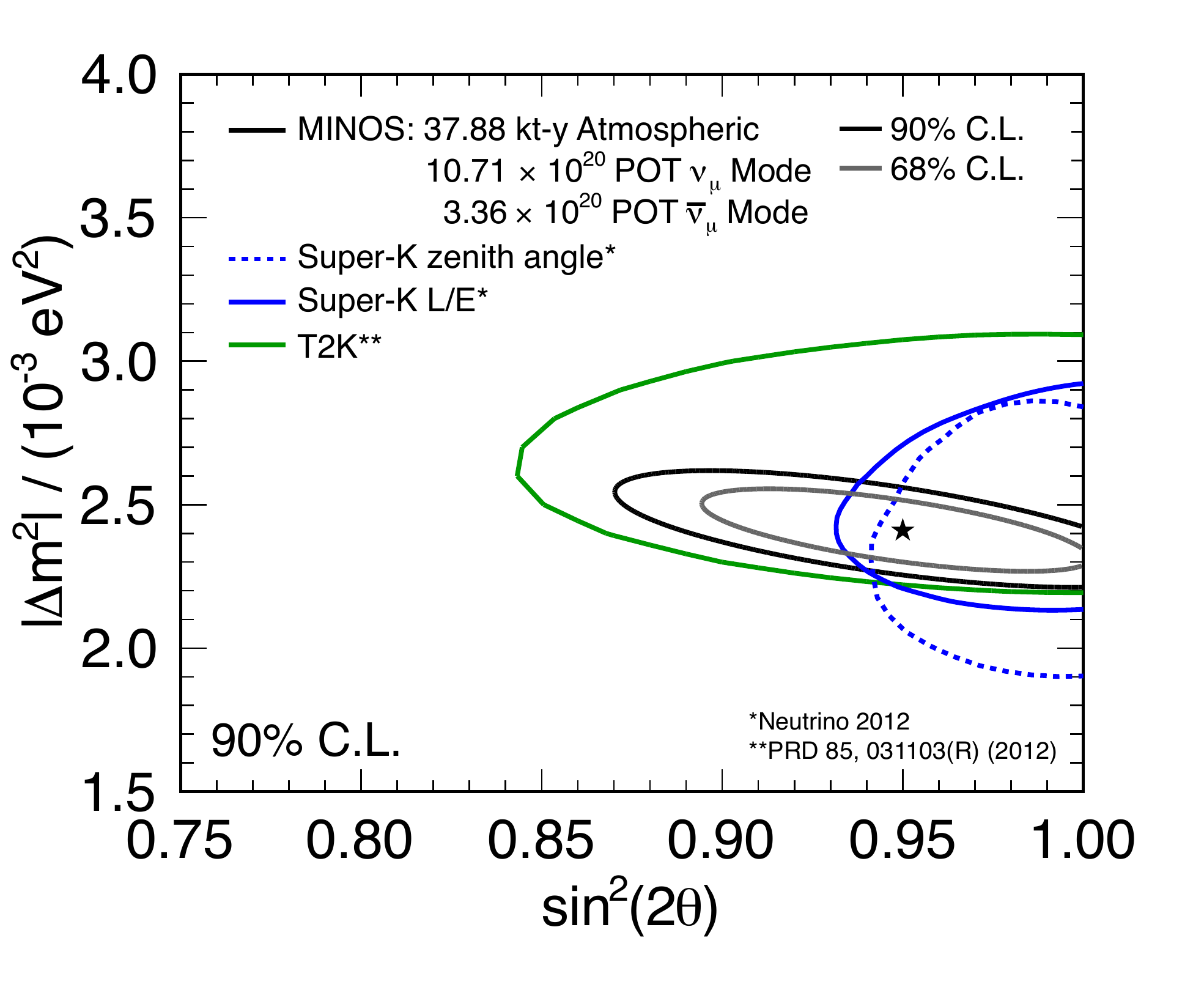}
\caption{The allowed regions for the atmospheric oscillation
  parameters $|\Delta m^{2}|$ and $\sin^{2}(2\theta)$, assuming
  identical neutrino and antineutrino oscillation parameters. The
  MINOS result is compared to measurements from
  Super-Kamiokande~\cite{ref:SuperKRecent} and T2K~\cite{ref:T2K}.}
\label{fig:CCContours}
\end{figure}

All the observed \numu and \numubar CC interactions are fit according
to equation~(\ref{eqn:NuMuDisappearance}), under the assumption that
neutrinos and antineutrinos have the same oscillation parameters. The
resulting measurement of $|\Delta m^{2}|$ and $\sin^{2}(2\theta)$ is
shown in Figure~\ref{fig:CCContours}. The fit yields $|\Delta m^{2}| =
\unit[(2.41^{+0.09}_{-0.10})\times 10^{-3}]{eV^{2}}$ and
$\sin^{2}(2\theta) = 0.950^{+0.035}_{-0.036}$, disfavouring maximal
mixing at the 86\% confidence level. Figure~\ref{fig:CCContours}
compares this measurement to those from
Super-Kamiokande~\cite{ref:SuperKRecent} and T2K~\cite{ref:T2K}. The
MINOS measurement is the most precise determination of $|\Delta
m^{2}|$, and all measurements of $\sin^{2}(2\theta)$ are consistent.

\section{Muon antineutrino disappearance}

In the standard model of neutrino oscillation, neutrinos and
antineutrinos obey the same parameters, with $\mathcal{CPT}$ symmetry
requiring that the masses of particles and antiparticles are
identical. The most sensitive test of this symmetry in other sectors
is from the kaon system~\cite{ref:PDGKaon}.  The data from the
antineutrino-enhanced beam enables the first direct comparison of the
neutrino and antineutrino oscillation parameters in the atmospheric
region. This comparison provides a limit on non-standard interactions
with the matter passed through by the neutrino
beam~\cite{ref:NSITheoryWolf,ref:NSITheoryValle,ref:NSITheoryGonz,ref:NSITheoryFried,ref:ZeynepThesis,ref:TonyNSI,ref:KoppFitsMINOSNSI}.

\begin{figure}
\centering
\includegraphics[width=0.7\textwidth]{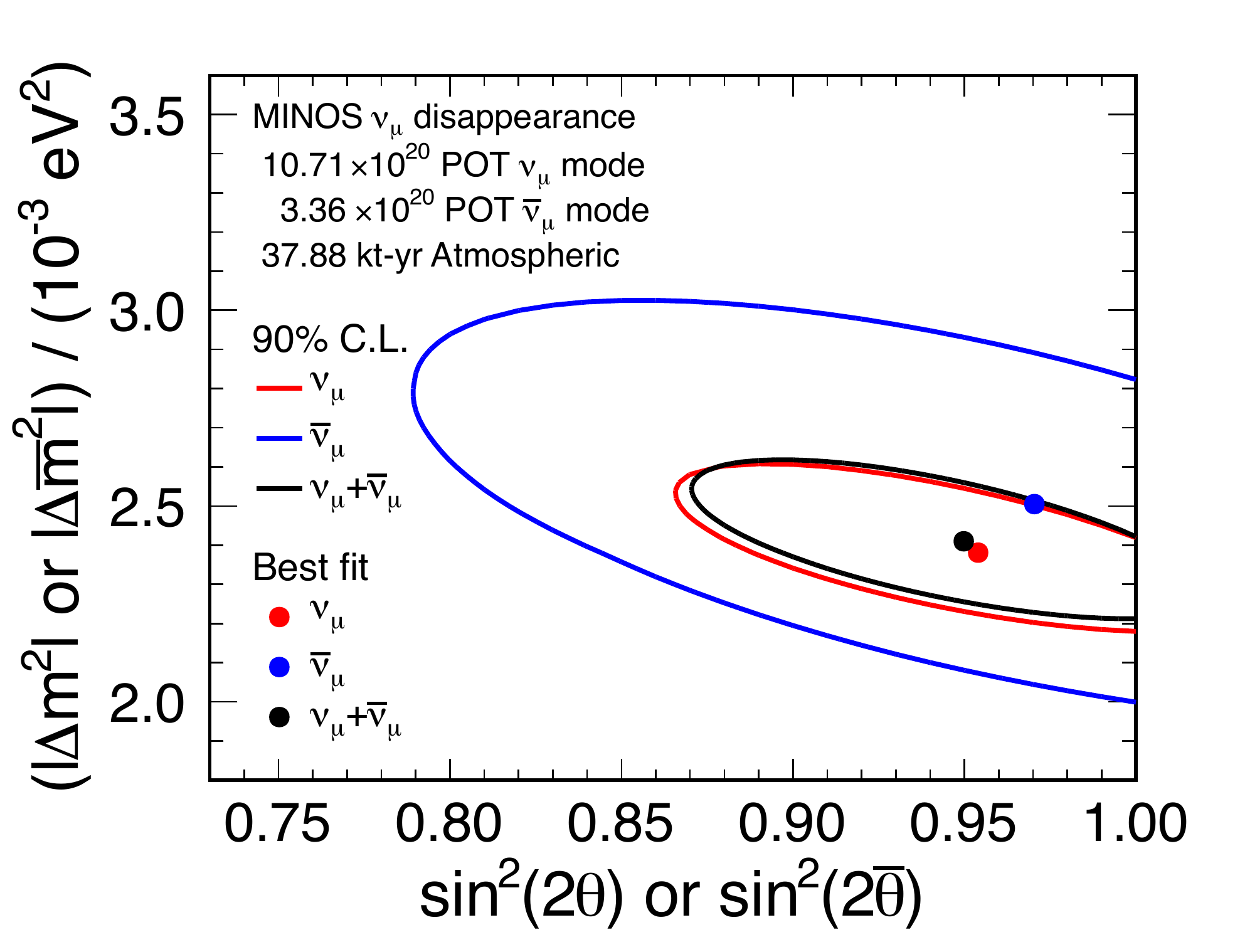}
\caption{The allowed region for antineutrino oscillation parameters
  (blue line), compared to the region measured with neutrinos alone
  (red line) and the region measured using both neutrinos and
  antineutrinos under the assumption they have the same parameters
  (black line).}
\label{fig:AntineutrinoContours}
\end{figure}

Figure~\ref{fig:CCFDSpectrum} showed the energy spectra of \numubar
interactions observed in the FD.  These spectra can be fit, allowing
the antineutrino oscillation parameters to differ from those for
neutrinos.  This fit yields the antineutrino parameter measurement
shown in Figure~\ref{fig:AntineutrinoContours}: $|\Delta
\overline{m}^{2}| = \unit[(2.50^{+0.23}_{-0.25})\times
  10^{-3}]{eV^{2}}$ and
$\sin^{2}(2\overline{\theta})=0.97^{+0.03}_{-0.08}$. This is in
excellent agreement with the parameters measured with neutrinos
neutrinos alone (the red line in
Figure~\ref{fig:AntineutrinoContours}).

\begin{figure}
\centering
\includegraphics[width=0.7\textwidth]{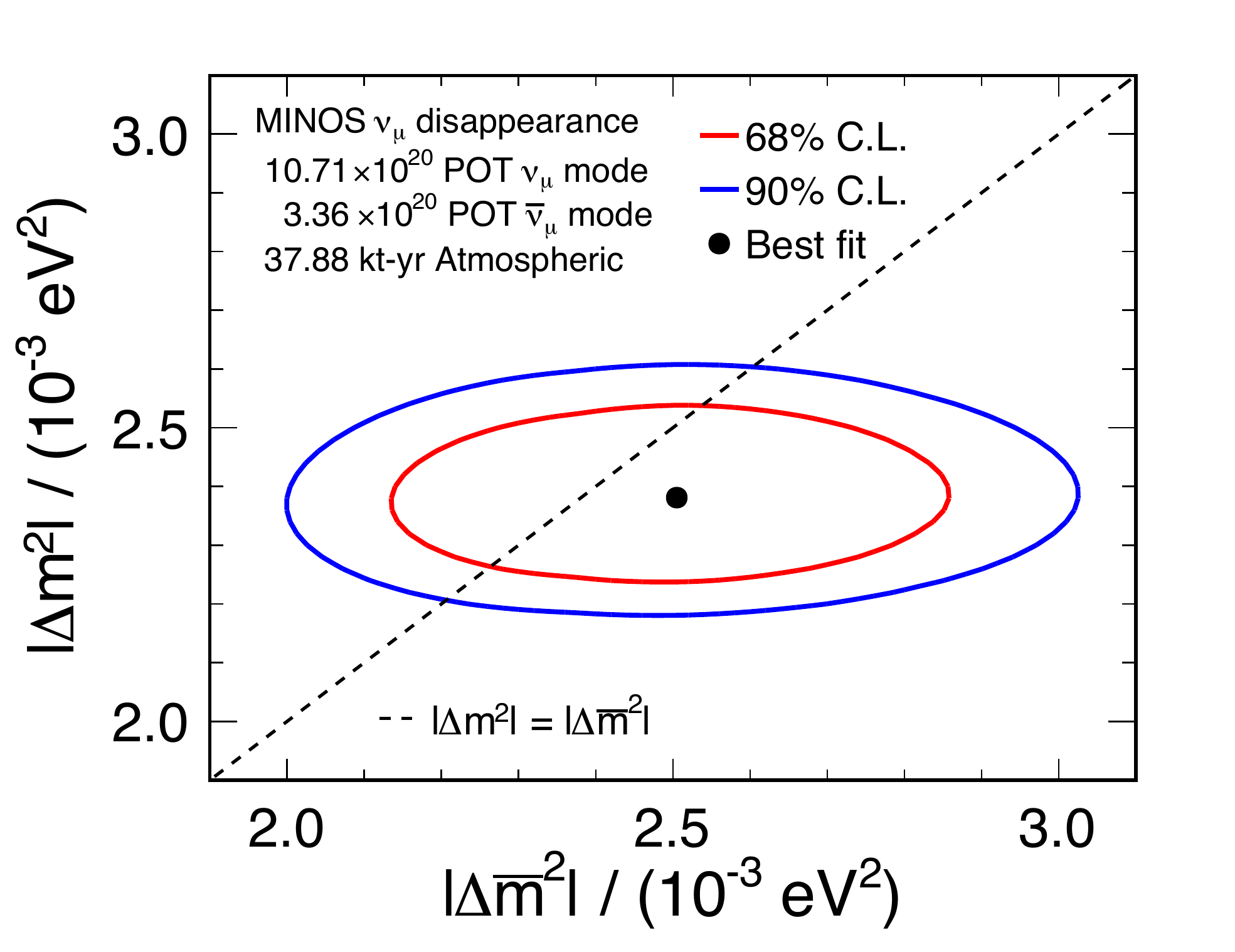}
\caption{The measured limit on the difference between the mass
  splittings of neutrinos and antineutrinos.}
\label{fig:NuNuBarDifference}
\end{figure}

The measured limit on the difference between the neutrino and
antineutrino mass splittings is shown in
Figure~\ref{fig:NuNuBarDifference}, and is $|\Delta
\overline{m}^{2}|-|\Delta m^{2}| =
\unit[(0.12^{+0.24}_{-0.26})\times10^{-3}]{eV^{2}}$.

\section{Electron neutrino appearance}

\begin{figure}
\centering
\includegraphics[width=0.32\textwidth]{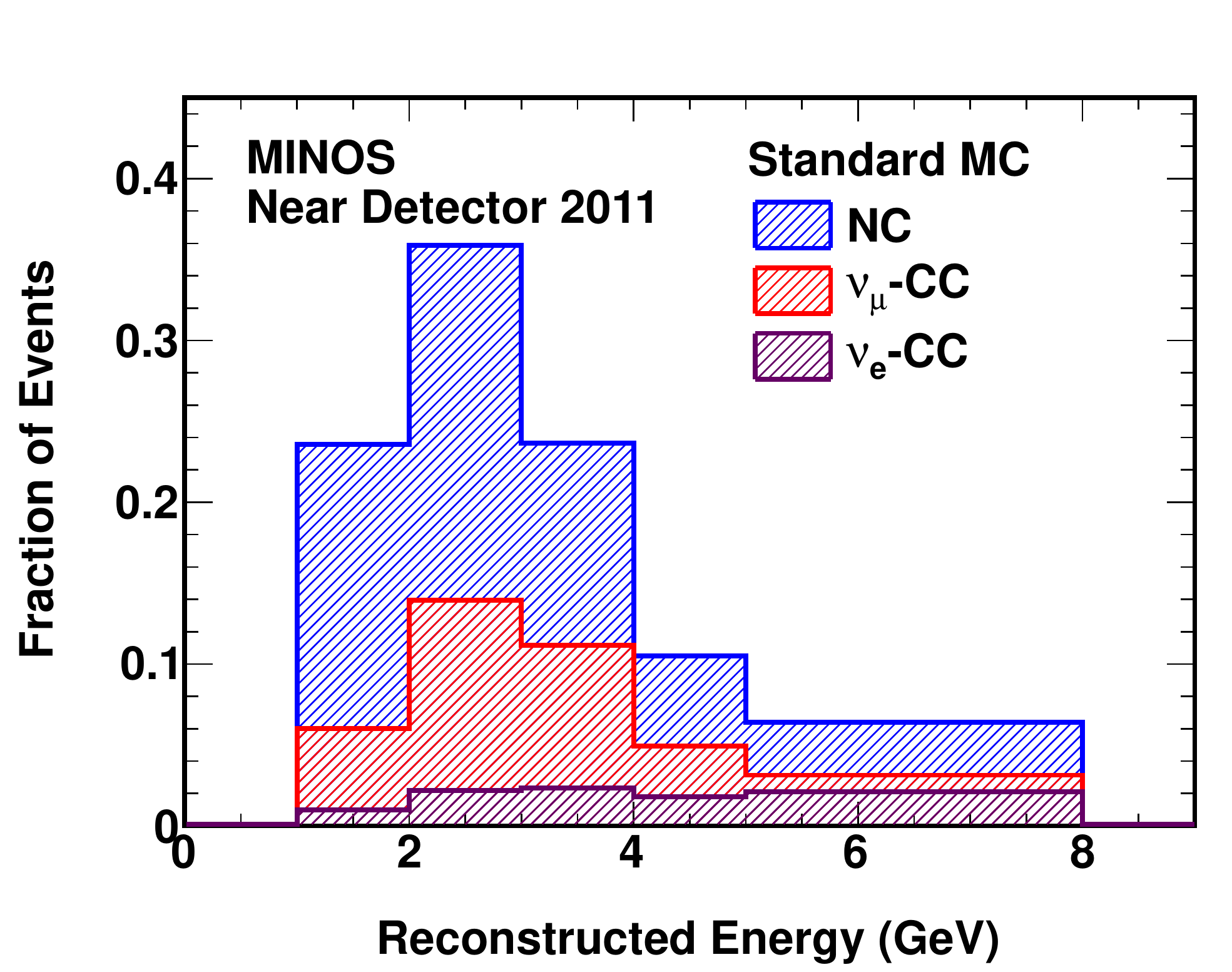}
\includegraphics[width=0.32\textwidth]{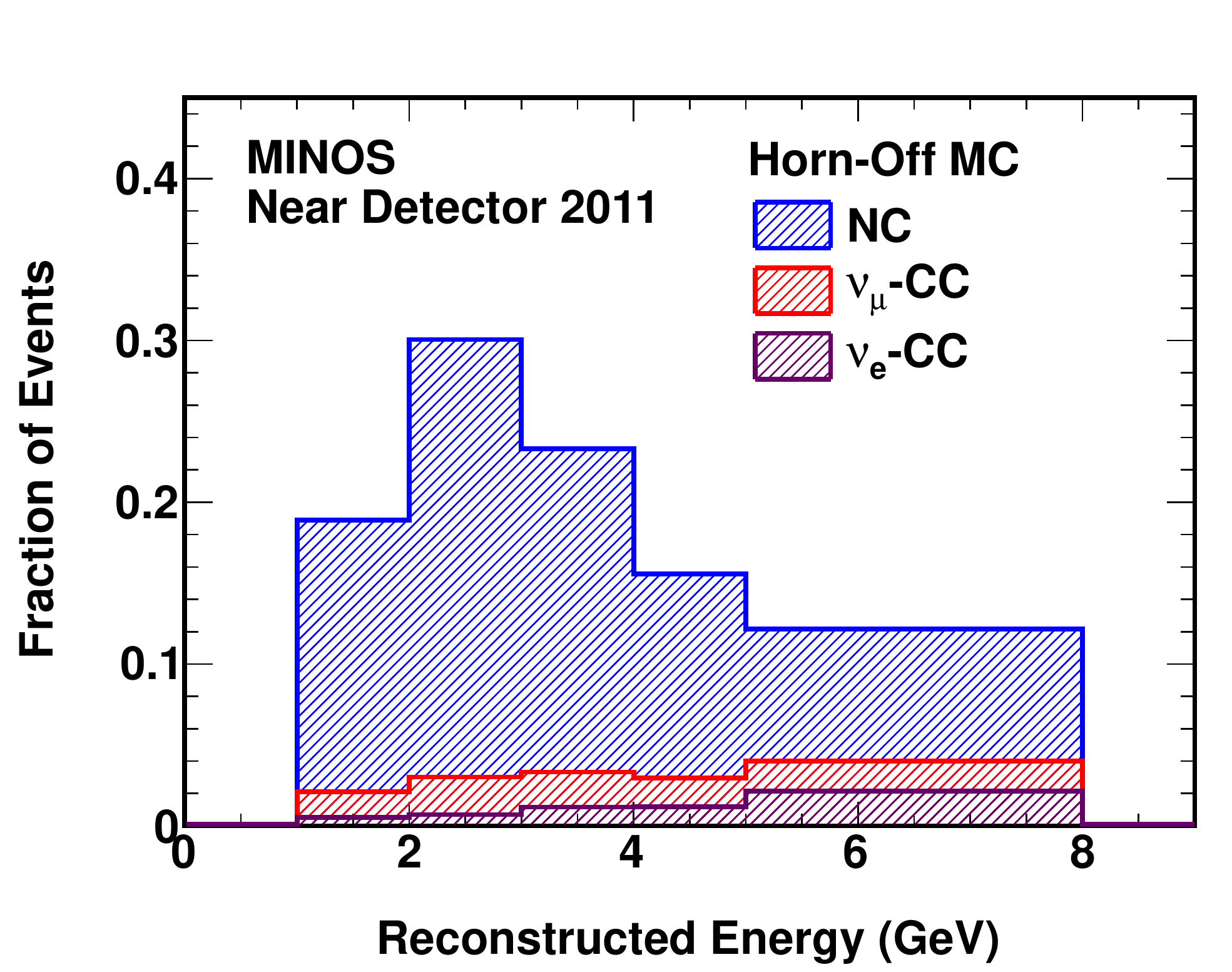}
\includegraphics[width=0.32\textwidth]{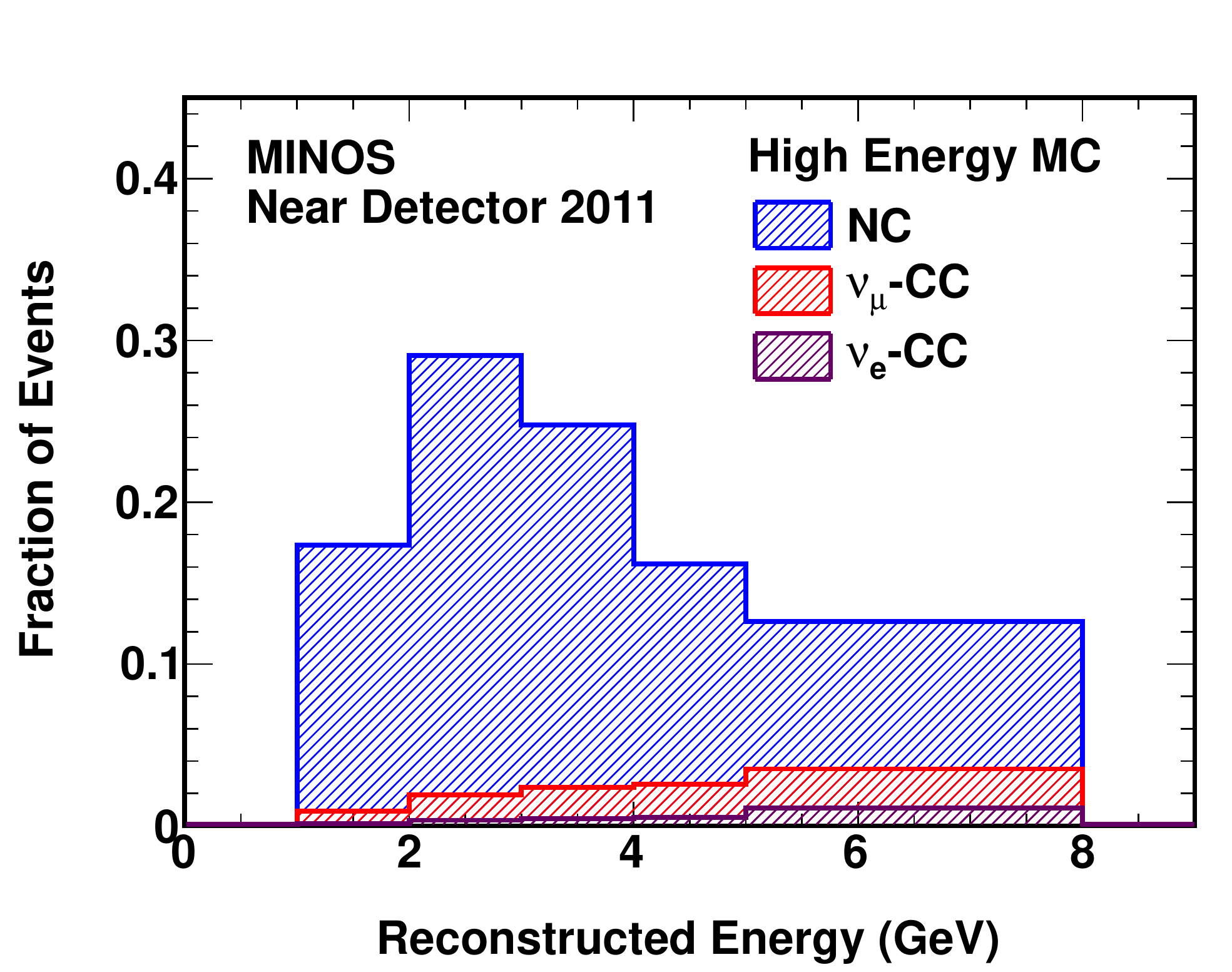}
\caption{The contribution of the three components to the background in
  the \nue appearance search, as simulated in the ND. Left: the
  regular beam data; middle: with no current in the NuMI focusing
  horns; right: a high energy beam configuration.}
\label{fig:NuEHoHo}
\end{figure}

A search for \nue and \nuebar appearance in the \numu and \numubar
beams enables a measurement of the mixing angle $\theta_{13}$.  It is
critical to know the level of background to the \nue sample in the FD.
The energy spectrum of background events measured in the ND is used to
predict the spectrum expected in the FD.  However, the background
consists of three components: NC interactions, CC \numu and \numubar
interactions, and the intrinsic \nue component in the beam.  The
relative contribution between the ND and FD is different for all of
these components, since they are affected differently by oscillation,
and the kinematics of the production in the beam are
different. Therefore each background must be individually measured.
The NuMI beam can be configured to produce neutrino beams of varying
energy, by altering the current passing through the magnetic horns and
changing the relative positions of the target and horns.  Between
these different beam configurations, the relative contributions of the
three background components changes in a well understood way, as shown
in Figure~\ref{fig:NuEHoHo}.  By comparing the ND data to the
simulation in the three different beam configurations shown in the
figure, the contributions of the three background components can be
extracted~\cite{ref:JoaoThesis}.

\begin{figure}
\centering
\includegraphics[width=0.7\textwidth]{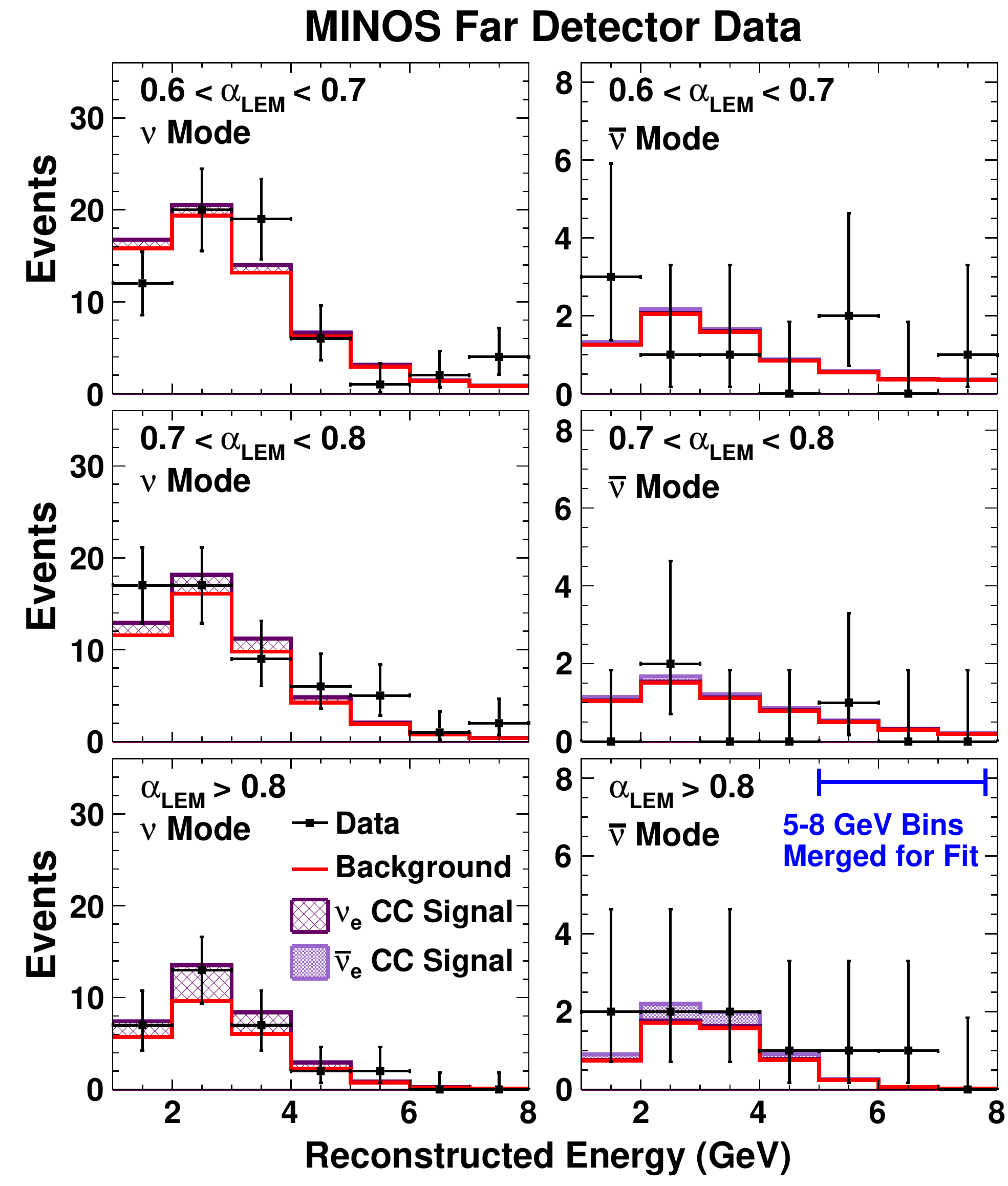}
\caption{The CC \nue (left) and \nuebar (right) candidate events
  selected in the FD, compared to the expectation without any \nue
  appearance (red) and with the best fit for $\theta_{13}$
  (purple). The events are divided into bins of the Library Event
  Matching discriminant variable.}
\label{fig:NuEDataSpectrum}
\end{figure}

Using the data-driven background extraction procedure, a total of
127.7 background events are expected at the FD in the
neutrino-dominated beam, and 17.5 events in the antineutrino-enhanced
beam.  In the data, 152 and 20 events are observed, respectively.
Figure~\ref{fig:NuEDataSpectrum} shows the energy spectra of these
events, divided into bins of the Library Event Matching discriminant
variable.

\begin{figure}
\centering
\includegraphics[width=0.6\textwidth]{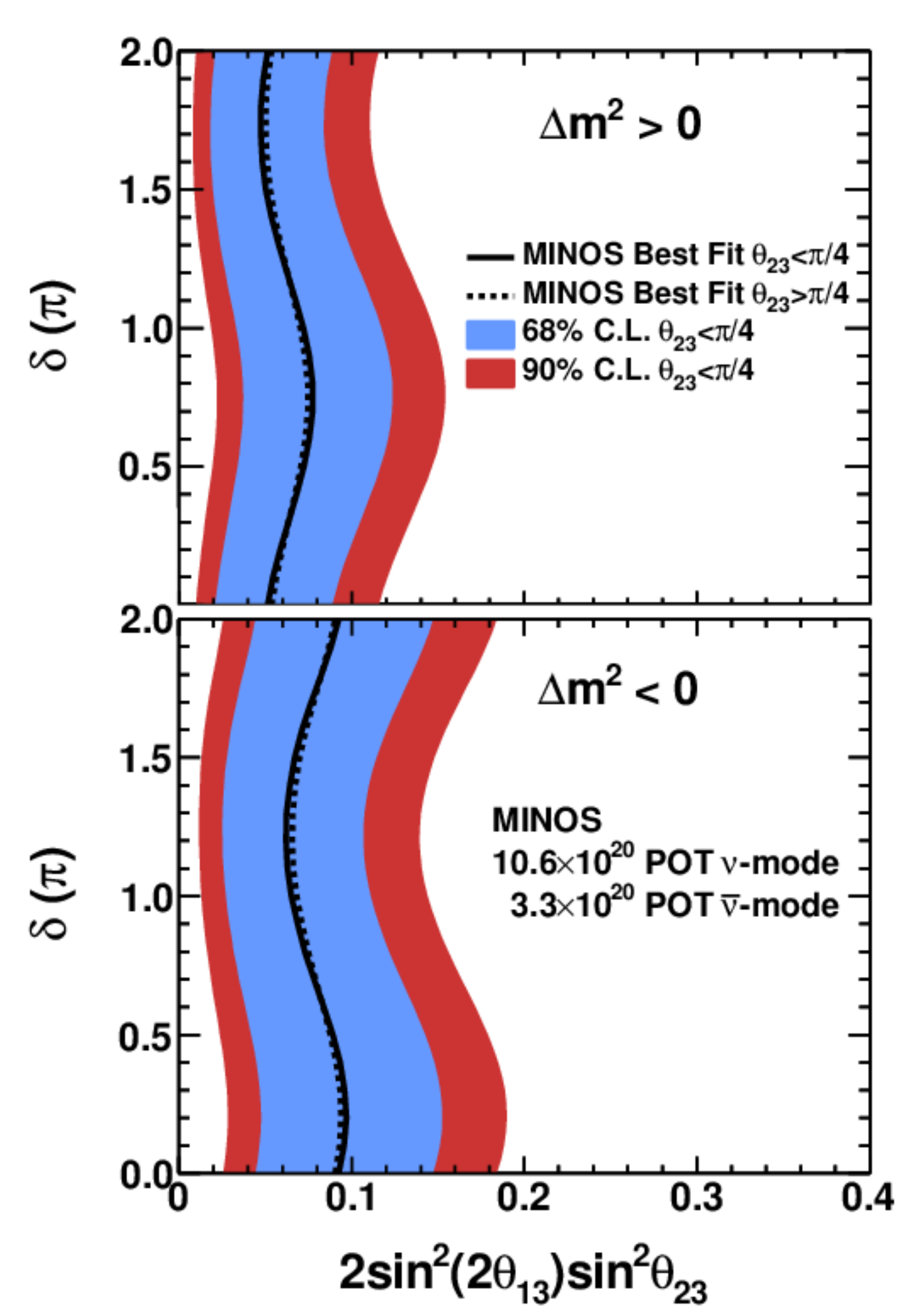}
\caption{The allowed regions for $2\sin^{2}(2\theta_{13})\sin^{2}(\theta_{23})$.}
\label{fig:NuEContour}
\end{figure}

The data are fit to extract a measurement of $\theta_{13}$.  The
resulting measurement is shown in Figure~\ref{fig:NuEContour}.  The
measured value of $\theta_{13}$ depends on the $\mathcal{CP}$
violating phase $\delta$, which directly affects the \nue and \nuebar
appearance probabilities, and the mass hierarchy, which affects the
appearance probabilities through the interactions of the neutrinos
with the matter in the Earth's crust. Assuming a normal mass
hierarchy, $\delta=0$ and $\theta_{23}<\pi/4$, MINOS measures
$2\sin^{2}(2\theta_{13})\sin^{2}(\theta_{23})=0.051^{+0.038}_{-0.030}$. Assuming
an inverted mass hierarchy, $\delta=0$ and $\theta_{23}<\pi/4$, MINOS
measures
$2\sin^{2}(2\theta_{13})\sin^{2}(\theta_{23})=0.093^{+0.054}_{-0.049}$.

\begin{figure}
\centering
\includegraphics[width=0.7\textwidth]{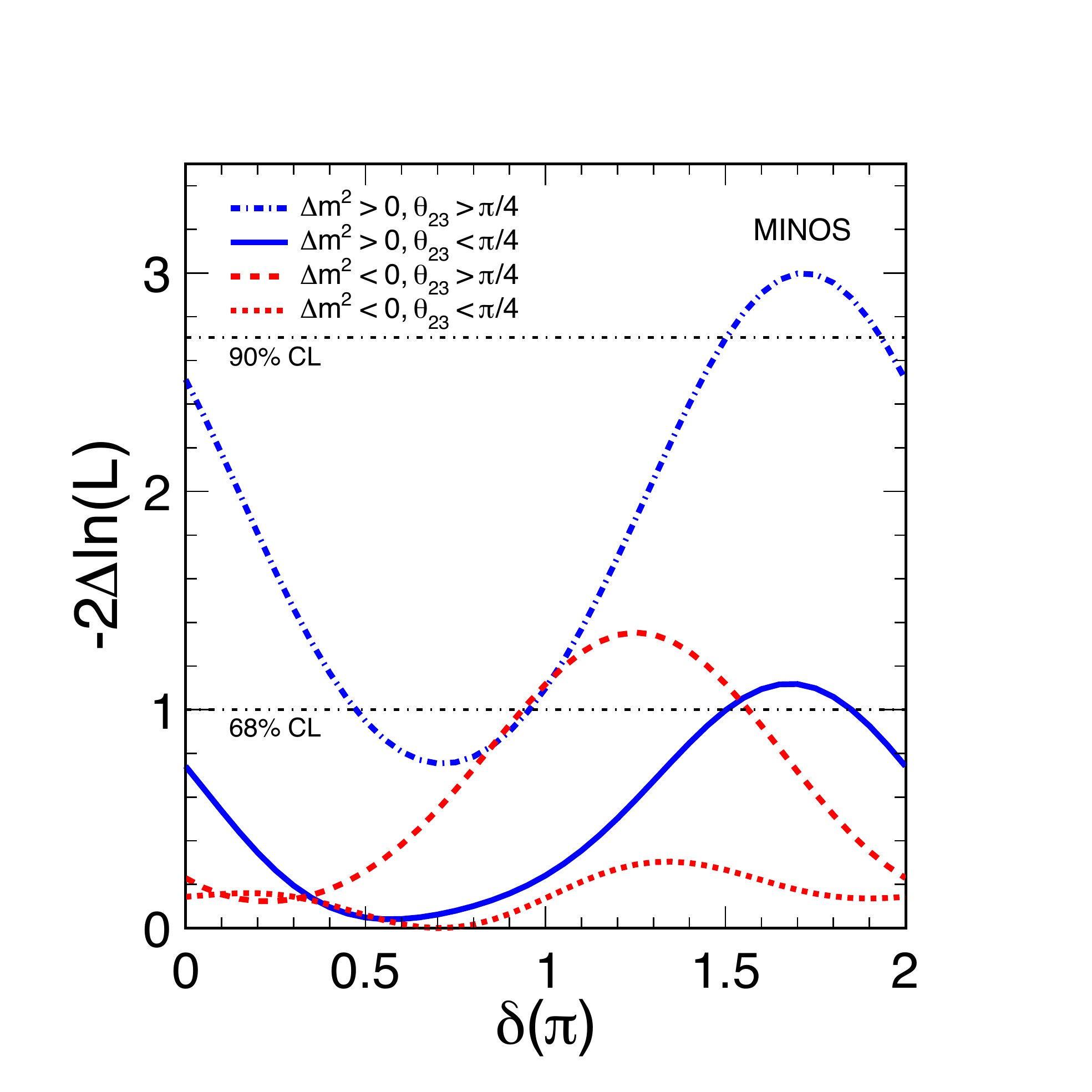}
\caption{The values of likelihood $L$ by which various values of the
  $\mathcal{CP}$ violating parameter $\delta$, the mass hierarchy, and the octant
  of $\theta_{23}$ are disfavoured.}
\label{fig:NuEDeltaHierarchy}
\end{figure}

This MINOS measurement is the first ever search for \nuebar appearance
in a long-baseline \numubar beam, and the first search for \nue and
\nuebar appearance with significant matter effects.  Both of these
effects provide some sensitivity to the neutrino mass hierarchy and
$\mathcal{CP}$ violation. The sensitivity of MINOS to these parameters
is modest, but this contributes the first analysis of the type that
will be used by all future long-baseline experiments.  The resulting
values of the likelihood by which MINOS disfavours various values of
these parameters are shown in Figure~\ref{fig:NuEDeltaHierarchy}~\cite{ref:SchreckenbergerThesis}.

\section{Neutral-current interaction rate}

\begin{figure}
\centering
\includegraphics[width=0.7\textwidth]{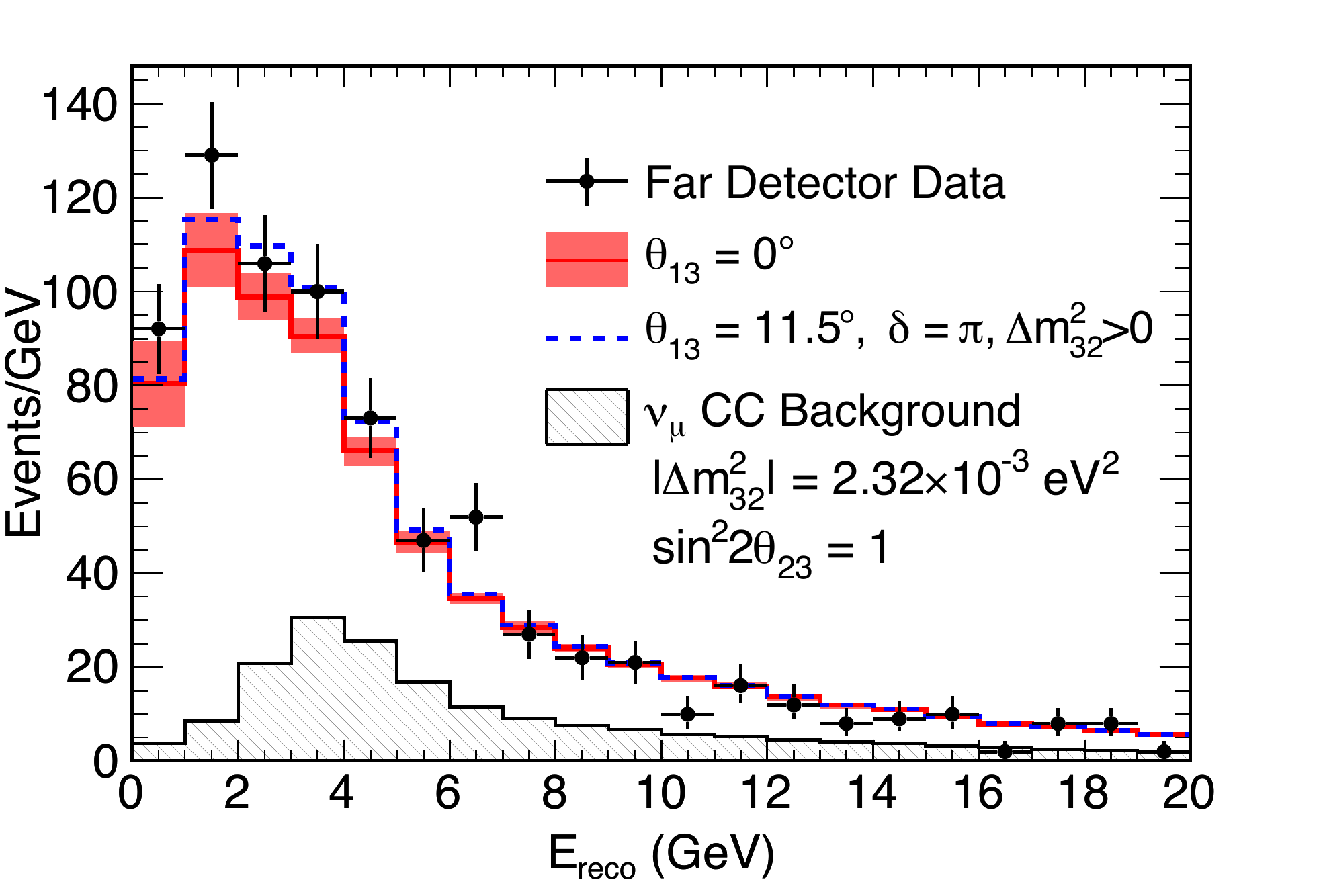}
\caption{The black dots show the energy spectrum of NC interactions
  observed in the Far Detector. The red show the expectation in the
  case of no sterile neutrinos and $\theta_{13}=0$; the blue dashed
  line shows the same expectation with $\theta_{13}=11.5^{\circ}$.}
\label{fig:NCFDSpectrum}
\end{figure}

The energy spectrum of NC interactions in the FD should be unchanged
by standard neutrino oscillation.  The existence of one or more
sterile neutrino flavours, $\nu_{s}$, could cause a deficit in the
observed NC interaction rate.  As with all the MINOS oscillation
analyses, the energy spectrum of NC interactions observed in the ND
(which was shown in Figure~\ref{fig:NCND}) is used to predict the
spectrum expected at the FD~\cite{ref:JasonThesis}.  The FD
expectation is shown in Figure~\ref{fig:NCFDSpectrum}, with the dashed
blue line taking into account \nue appearance corresponding to
$\theta_{13}=11.5^{\circ}$ (at the limit set by
CHOOZ~\cite{ref:CHOOZ}, and a little above the current accepted
value~\cite{ref:DayaBay,ref:Reno,ref:DoubleChooz,ref:T2KTheta13,ref:MINOSNuE2012}).
The data are also shown in the figure, and are in good agreement with
the expectation, confirming the standard model of neutrino
oscillation.

The limit on the coupling of sterile to active neutrinos can be
quantified by defining $f_{s}$, the fraction of disappearing \numu which have
oscillated into $\nu_{s}$:
\begin{equation}
f_{s}=\frac{P_{\numu\rightarrow\nu_{s}}}{1-P_{\numu\rightarrow\numu}}.
\end{equation}
Assuming $\theta_{13}=11.5^{\circ}$, MINOS limits $f_{s}<0.40$ at the
90\% confidence limit.

\section{The Future: MINOS+}

\begin{figure}
\centering
\includegraphics[width=0.7\textwidth]{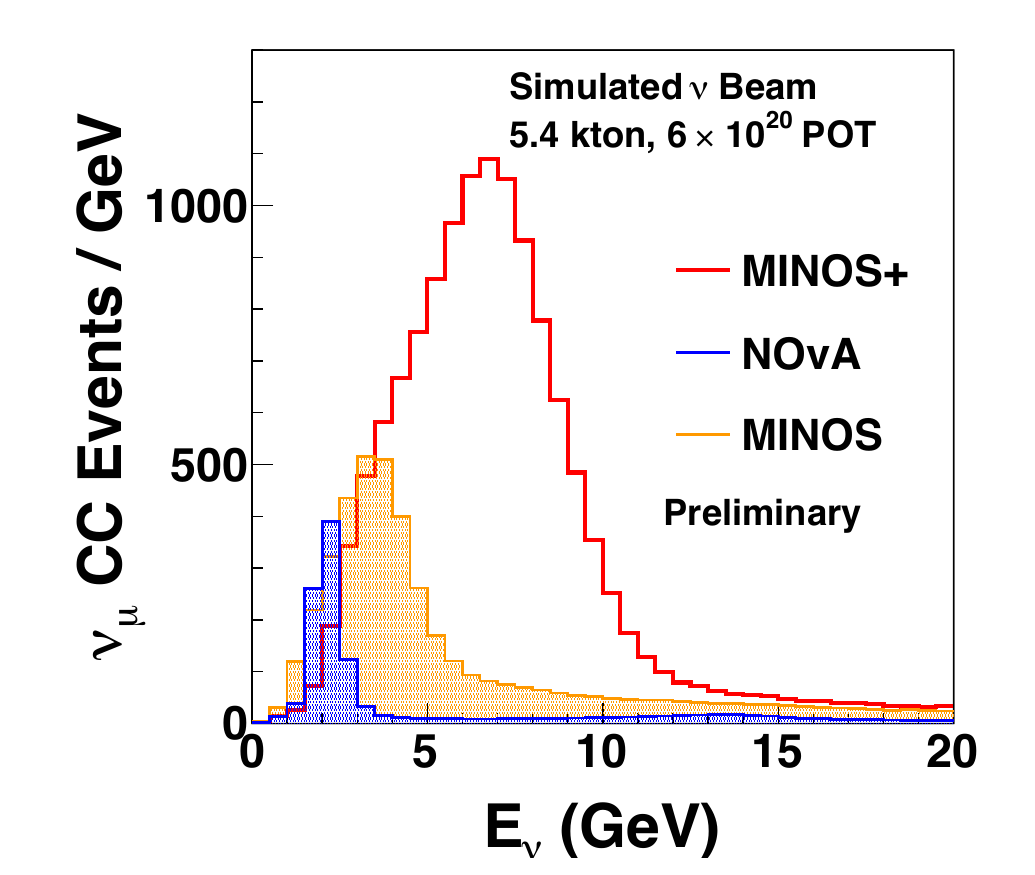}
\caption{The \numu energy spectrum that will be observed by the
  MINOS+ FD, compared to the spectra observed by MINOS and NO$\nu$A.}
\label{fig:MINOSPlusSpectrum}
\end{figure}

The MINOS experiment has made some very important contributions to our
understanding of neutrino oscillation physics, and has finished taking
data with the beam for which it was designed.  However, the experiment
will continue taking data and producing new results for the next few
years as MINOS+~\cite{ref:MINOSPlusProposal}. The NuMI beam is being
upgraded to a higher energy and intensity for the NO$\nu$A experiment,
the far detector of which will sit \unit[14]{mrad} off-axis.  NO$\nu$A
will receive a narrow-band beam, peaking at around \unit[2]{GeV},
which is ideal for searching for \nue appearance since the background
seen in MINOS from NC interactions of high energy neutrinos will be
heavily reduced.  Figure~\ref{fig:MINOSPlusSpectrum} shows that the
MINOS FD will see an intense \numu beam, peaking at around
\unit[7]{GeV}.  In this configuration, MINOS+ will observe around
4,000 CC \numu interactions in the FD each year; unprecedented
statistics for a long-baseline oscillation experiment.  This will
offer a unique, high precision test of the three-flavour oscillation
paradigm.

\begin{figure}
\centering
\includegraphics[width=0.7\textwidth]{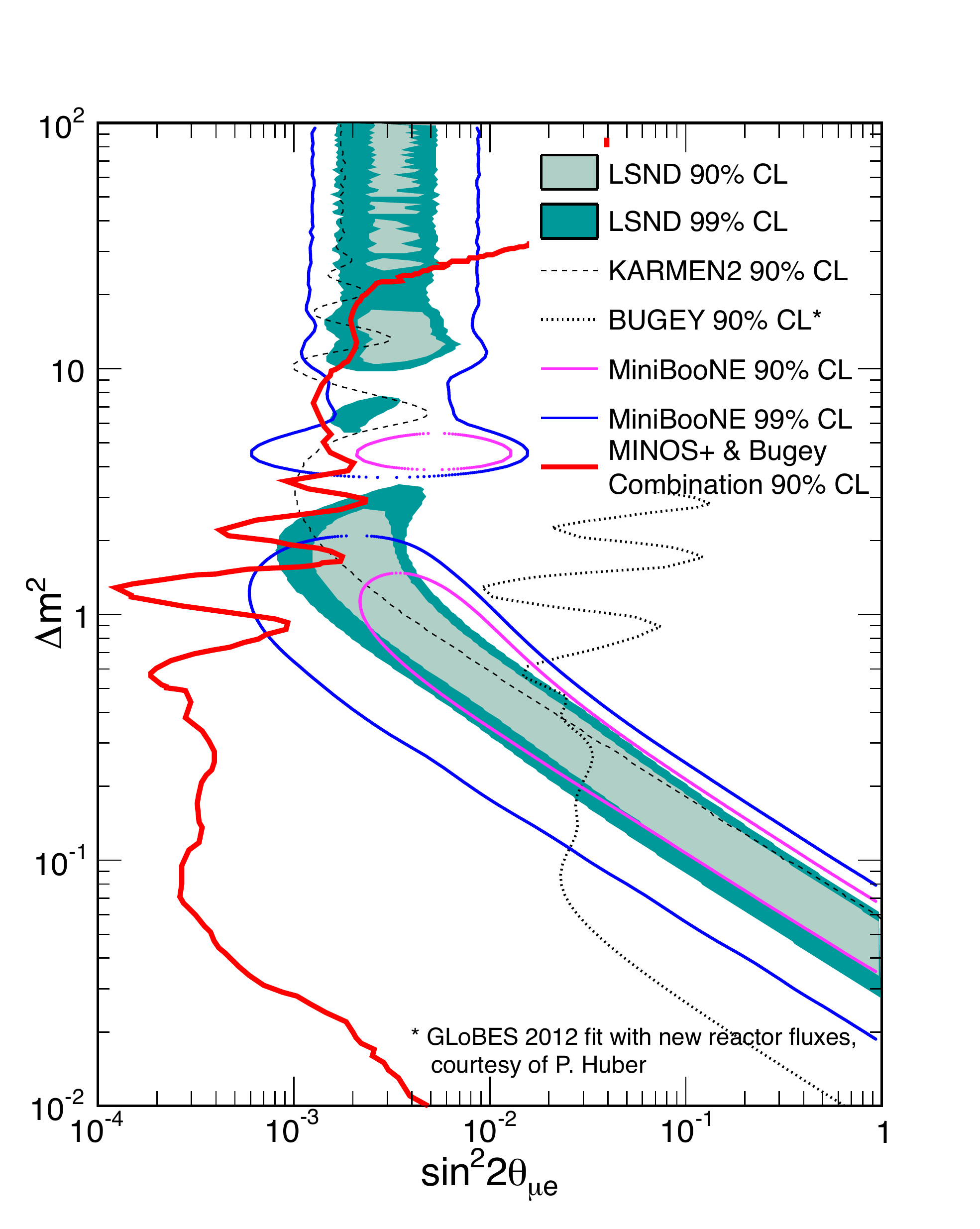}
\caption{The sensitivity of MINOS+ to the existence of sterile
  neutrinos, when combined with data from the Bugey~\cite{ref:Bugey}
  reactor neutrino experiment. $\Delta m^{2}$ is the splitting between
  the three known neutrino mass states and a new, fourth
  state. $\theta_{\mu e}$ is the mixing angle governing
  $\numu\rightarrow\nue$ transitions when a fourth, sterile neutrino
  state is introduced into the PMNS mixing matrix.}
\label{fig:MINOSPlusBugeyCombo}
\end{figure}

MINOS+ will be able to make a very sensitive search for the sterile
neutrinos suggested by the LSND~\cite{ref:LSND} and
MiniBooNE~\cite{ref:MiniBooNE} data, and by some interpretations of
reactor neutrino data~\cite{ref:ReactorNeutrinoAnomaly}.  This search
will cover more than three orders of magnitude in the mass splitting
between the sterile and active neutrinos.  The sensitivity of the
MINOS+ experiment, when combined with the Bugey reactor neutrino
data~\cite{ref:Bugey}, is shown in
Figure~\ref{fig:MINOSPlusBugeyCombo}; MINOS+ has the potential to rule
out much of the LSND allowed region.

\section{Conclusion}

The MINOS experiment was conceived at a time when neutrino oscillation
had only recently been confirmed as the solution to the problem of
neutrino flavour change.  It has played a hugely influential role in
bringing neutrino oscillation physics into an era of precision
measurement.  MINOS's measurement of the largest neutrino mass
splitting is the most precise in the world.  MINOS has made the first
direct precision measurement of the corresponding antineutrino
parameters, a measurement that promises to remain the world's most
precise for many years.  And MINOS has played a role in the discovery
of a non-zero value for $\theta_{13}$.

Now that the value of $\theta_{13}$ is known, the neutrino physics
community can move on to determine the neutrino mass hierarchy, and to
search for $\mathcal{CP}$ violation in the neutrino sector.  MINOS has
pioneered a number of techniques that will be used by all future
experiments.  The two-detector setup, all important in reducing the
impact of systematic uncertainties, is the design of choice for any
new experiment; and MINOS has demonstrated methods of using a near
detector to determine the expectation at a far detector.  MINOS has
performed the first search for \nuebar appearance in a \numubar beam,
and the first search for \nue and \nuebar appearance with significant
matter effects, demonstrating the analysis techniques that will be
used to determine the mass hierarchy and $\mathcal{CP}$ violation
parameter.

In the second half of 2013, MINOS will begin taking data as the MINOS+
experiment, which will make ever more precise tests of the
three-flavour neutrino oscillation paradigm and set world-leading
limits on the existence of sterile neutrinos.  This is an exciting
future for an experiment that, with a decade of data taking so far,
has already created a lasting legacy for itself in our understanding
of the neutrino.

\acknowledgments

The work of the MINOS and MINOS+ collaborations is supported by the US
DoE, the UK STFC, the US NSF, the State and University of Minnesota,
the University of Athens in Greece, and Brazil's FAPESP and CNPq. We
are grateful to the Minnesota Department of Natural Resources, the
crew of the Soudan Underground Laboratory, and the personnel of
Fermilab, for their vital contributions.


\begin{thebibliography}{99}
  
\bibitem{ref:StanNeutrino1998} S. Wojcicki,
  Nucl.\ Phys.\ Proc.\ Suppl.\ {\bf 77}, 182 (1999), proceedings of
  the XVIII International Conference on Neutrino Physics and
  Astrophysics (Neutrino 1998), Takayama, Japan, June 1998.
  
\bibitem{ref:Homestake} R. Davis, D.S. Harmer and K.C. Hoffman,
  Phys.\ Rev.\ Lett.\ {\bf 20}, 1205 (1968).
  
\bibitem{ref:Sage} A.I. Abazov {\it et al.} (SAGE),
  Phys.\ Rev.\ Lett.\ {\bf 67}, 3332 (1991).
\bibitem{ref:Gallex}P. Anselmann {\it et al.} (GALLEX),
  Phys. Lett. {\bf B285}, 376 (1992).
  
\bibitem{ref:Nusex} M. Aglietta {\it et al.}  (NUSEX),
  Europhys.\ Lett.\ {\bf 8}, 611 (1989).
\bibitem{ref:Kamiokande} K. Hirata {\it et al.} (Kamiokande),
  Phys. Lett. {\bf B280}, 146 (1992).
\bibitem{ref:IMB} R. Becker-Szendy {\it et al.}  (IMB),
  Phys. Rev. {\bf D46}, 3720 (1992).
\bibitem{ref:Frejus} K. Daum {\it et al.}  (Fr\'ejus), Z. Phys. {\bf
  C66}, 417 (1995).
\bibitem{ref:MACRO} S. Ahlen {\it et al.}  (MACRO) Phys.\ Lett.\ {\bf
  B357}, 481 (1995).
\bibitem{ref:Soudan2} W. Allison {\it et al.}  (Soudan2),
  Phys.\ Lett.\ {\bf B391}, 491 (1997).
  
\bibitem{ref:SuperKFirst} Y. Fukuda {\it et al.} (Super-Kamiokande),
  Phys.\ Rev.\ Lett.\ {\bf 81}, 1562 (1998).
  
\bibitem{ref:SNO} Q.R. Ahmad {\it et al.} (SNO),
  Phys.\ Rev.\ Lett.\ {\bf 87}, 071301 (2001);
  Phys.\ Rev.\ Lett.\ {\bf 89}, 011301 (2002).
  
\bibitem{ref:PMNS} B. Pontecorvo, JETP {\bf 34}, 172 (1958); V.N. Gribov and B. Pontecorvo, Phys.\ Lett.\ {\bf B28}, 493 (1969); Z. Maki, M. Nakagawa and S. Sakata, Prog.\ Theor.\ Phys.\ {\bf 28}, 870 (1962).
  
\bibitem{ref:MINOSRHCFirst} P. Adamson {\it et al.} (MINOS), Phys.\ Rev.\ Lett.\ {\bf 107}, 021801 (2011).
\bibitem{ref:MINOSFHCNuBars} P. Adamson {\it et al.} (MINOS), Phys.\ Rev.\ {\bf D84}, 071103 (2011).
\bibitem{ref:MINOSRHCSecond} P. Adamson {\it et al.} (MINOS), Phys.\ Rev.\ Lett.\ {\bf 108}, 191801 (2012).
\bibitem{ref:MINOSCC2012} P. Adamson {\it et al.} (MINOS), Phys.\ Rev.\ Lett.\ {\bf 110}, 251801 (2013).
  
\bibitem{ref:NuMI} K. Anderson {\it et al.}, FERMILAB-DESIGN-1998-01 (1998).
  
\bibitem{ref:MINOSCC2006} D.G. Michael {\it et al.} (MINOS), Phys.\ Rev.\ Lett.\ {\bf 97}, 191801 (2006).
\bibitem{ref:MINOSCCPRD} P. Adamson {\it et al.} (MINOS), Phys.\ Rev.\ {\bf D77}, 072002 (2008).
\bibitem{ref:MINOSCC2008} P. Adamson {\it et al.} (MINOS), Phys.\ Rev.\ Lett.\ {\bf 101}, 131802 (2008).
\bibitem{ref:MINOSCC2010} P. Adamson {\it et al.} (MINOS), Phys.\ Rev.\ Lett.\ {\bf 106}, 181801 (2011).
  
\bibitem{ref:MINOSNuEFirst} P. Adamson {\it et al.} (MINOS), Phys.\ Rev.\ Lett.\ {\bf 103}, 261802 (2009).
\bibitem{ref:MINOSNuEPRD} P. Adamson {\it et al.} (MINOS), Phys.\ Rev.\ {\bf D82}, 051102 (2010).
\bibitem{ref:MINOSNuE2011} P. Adamson {\it et al.} (MINOS), Phys.\ Rev.\ Lett.\ {\bf 107}, 181802 (2011).
\bibitem{ref:MINOSNuE2012} P. Adamson {\it et al.} (MINOS), Phys.\ Rev.\ Lett.\ {\bf 110}, 171801 (2013).
  
\bibitem{ref:MINOSNCFirst} P. Adamson {\it et al.} (MINOS), Phys.\ Rev.\ Lett.\ {\bf 101}, 221804 (2008).
\bibitem{ref:MINOSNCPRD} P. Adamson {\it et al.} (MINOS), Phys.\ Rev.\ {\bf D81}, 052004 (2010).
\bibitem{ref:MINOSNC2011} P. Adamson {\it et al.} (MINOS), Phys.\ Rev.\ Lett.\ {\bf 107}, 011802 (2011).
  
\bibitem{ref:ZarkoThesis} \v{Z}. Pavlovi\'c, PhD thesis, University of Texas at Austin (2008).
\bibitem{ref:Fluka} F. Ballarini {\it et al.} (Fluka), J. Phys.\ Conf.\ Ser.\ {\bf 41}, 151 (2006).
\bibitem{ref:GEANT4} S. Agostinelli {\it et al.} (GEANT4), Nucl.\ Instrum.\ Meth.\ {\bf A506}, 250 (2003).
\bibitem{ref:Flugg} G. Battistoni {\it et al.} (Flugg), AIP Conf.\ Proc.\ {\bf 896}, 31 (2007).

\bibitem{ref:MINOSNIM} D. Michael {\it et al.} (MINOS), Nucl.\ Instrum.\ Meth.\ {\bf A596}, 190 (2008).  

\bibitem{ref:MINOSAtmosFirst} P. Adamson {\it et al.} (MINOS), Phys.\ Rev.\ {\bf D73}, 072002 (2006).
\bibitem{ref:MINOSAtmosChargeSeparated} P. Adamson {\it et al.} (MINOS), Phys.\ Rev.\ {\bf D75}, 092003 (2007).
\bibitem{ref:MINOSAtmosPRD} P. Adamson {\it et al.} (MINOS), Phys.\ Rev.\ {\bf D86}, 052007 (2012).

\bibitem{ref:MikeThesis} M.A. Kordosky, PhD thesis, University of Texas at Austin (2004).
\bibitem{ref:TriciaThesis} P.L. Vahle, PhD thesis, University of Texas at Austin (2004).
\bibitem{ref:BackhouseThesis} C. Backhouse, DPhil thesis, University of Oxford (2011).
\bibitem{ref:kNN} T.M. Cover and P.E. Hart, IEE Trans.\ Inform.\ Theory {\bf 13}, 21 (1967).
\bibitem{ref:RustemThesis} R. Ospanov, PhD thesis, University of Texas at Austin (2008).
\bibitem{ref:JohnMarshallThesis} J.S. Marshall, PhD thesis, University of Cambridge (2008).
\bibitem{ref:PedroThesis} J.P. Ochoa, PhD thesis, Caltech (2009).
\bibitem{ref:RuthThesis} R. Toner, PhD thesis, University of Cambridge (2011).
\bibitem{ref:AnnaThesis} A. Holin, PhD thesis, University College London (2010).
\bibitem{ref:JoshThesis} J. Boehm, PhD thesis, Harvard University (2009).
\bibitem{ref:GemmaThesis} G. Tinti, DPhil thesis, University of Oxford (2010).
\bibitem{ref:MyThesis} J.J. Evans, DPhil thesis, University of Oxford (2008).
\bibitem{ref:StephenThesis} S.J. Coleman, PhD thesis, College of William \& Mary (2011).
\bibitem{ref:JessThesis} J.S. Mitchell, PhD thesis, University of Cambridge (2011).
\bibitem{ref:AaronThesis} A. McGowan PhD thesis, University of Minnesota (2007).
\bibitem{ref:StraitThesis} M. Strait, PhD thesis, University of Minnesota (2010).
\bibitem{ref:SuperKRecent} Y. Itow, in the proceedings of the XXV
  International Conference on Neutrino Physics and Astrophysics
  (Neutrino 2012), Kyoto, Japan, June 2012, to be published in
  Nucl.\ Phys.\ {\bf B}.
\bibitem{ref:T2K} K. Abe {\it et al.} (T2K), Phys.\ Rev.\ {\bf D85}, 031103 (2012).
\bibitem{ref:PDGKaon} K. Nakamura {\it et al.} (Particle Data Group),
  J. Phys.\ {\bf G37}, 075021 (2010). See pages 89 and 759.
\bibitem{ref:NSITheoryWolf} L. Wolfenstein, Phys.\ Rev.\ {\bf D17}, 2369 (1978).
\bibitem{ref:NSITheoryValle} J.W.F. Valle, Phys.\ Lett.\ {\bf B199}, 432 (1987).
\bibitem{ref:NSITheoryGonz} M.C. Gonzalez-Garcia {\it et al.}, Phys.\ Rev.\ Lett.\ {\bf 82}, 3202 (1999);
\bibitem{ref:NSITheoryFried} A. Friedland, C. Lunardini and M. Maltoni, Phys.\ Rev.\ {\bf D70}, 111301 (2004).
\bibitem{ref:ZeynepThesis} Z. Isvan, PhD thesis, University of Pittsburgh (2012).
\bibitem{ref:TonyNSI} W.A. Mann {\it et al.}, Phys.\ Rev.\ {\bf D82}, 113010
  (2010).
\bibitem{ref:KoppFitsMINOSNSI} J. Kopp {\it et al.}, Phys.\ Rev.\ {\bf D82}, 113002 (2010).
\bibitem{ref:JoaoThesis} J.A.B. Coelho, PhD thesis, Universidade Estadual de Campinas (2012).
\bibitem{ref:SchreckenbergerThesis} A. Schreckenberger, PhD thesis, University of Minnesota (2013).
\bibitem{ref:JasonThesis} D.J. Koskinen, PhD thesis, University College London (2009).
\bibitem{ref:CHOOZ} M. Apollonio {\it et al.} (CHOOZ), Eur.\ Phys.\ J. {\bf C27}, 331 (2003).
\bibitem{ref:DayaBay} F.P. An {\it et al.} (Daya Bay),
  Chin.\ Phys.\ {\bf C37}, 011001 (2013).
\bibitem{ref:Reno} S-B Kim {\it et al.} (RENO),
  Phys.\ Rev.\ Lett.\ {\bf 108}, 191802 (2012).
\bibitem{ref:DoubleChooz} Y. Abe {\it et al.}
  (Double Chooz), Phys.\ Rev.\ {\bf D86}, 052008 (2012).
\bibitem{ref:T2KTheta13} K. Abe {\it et al.}, arXiv:1304.0841 (2013).
\bibitem{ref:MINOSPlusProposal} G. Tzanakos {\it et al.}, FERMILAB-PROPOSAL-1016 (2011).
\bibitem{ref:LSND} A. Aguilar {\it et al.} (LSND), Phys.\ Rev.\ {\bf D64}, 112007 (2001).
\bibitem{ref:MiniBooNE} A.A. Aguilar-Arevalo {\it et al.} (MiniBooNE), Phys.\ Rev.\ Lett.\ {\bf 103}, 111801 (2009); Phys.\ Rev.\ Lett.\ {\bf 110}, 161801 (2013).
\bibitem{ref:ReactorNeutrinoAnomaly} G. Mention {\it et al.}, Phys.\ Rev.\ {\bf D83}, 073006 (2011).
\bibitem{ref:Bugey} Nucl.\ Phys.\ {\bf B434}, 503 (1995). The Bugey
  limit has been recomputed by P. Huber, accounting for the new
  reactor flux calculations used in~\cite{ref:ReactorNeutrinoAnomaly}.
  
\end{thebibliography}
\end{document}